\documentclass[journal]{IEEEtran}
\usepackage{amsmath,amsfonts,mathtools,amsthm,amssymb,dsfont}
\usepackage{graphicx}
\usepackage{algorithmic,algorithm}
\usepackage{amsthm}
\usepackage{subfigure,color,url}
\usepackage{epstopdf}
\usepackage[noadjust]{cite}

\def \bangle{ \atopwithdelims \langle \rangle}

\newtheorem{theorem}{Theorem}
\newtheorem{lemma}{Lemma}
\newtheorem{prop}{Proposition}
\newtheorem{corollary}{Corollary}

\newcommand\relphantom[1]{\mathrel{\phantom{#1}}}%用于模拟某个字符的长度

\ifCLASSINFOpdf
\else
\fi

\begin{document}
	\title{Coverage Analysis for Millimeter Wave Networks: The Impact of Directional Antenna Arrays}
	
	\author{Xianghao~Yu,~\IEEEmembership{Student Member,~IEEE},
		Jun~Zhang,~\IEEEmembership{Senior Member,~IEEE},
		Martin~Haenggi,~\IEEEmembership{Fellow,~IEEE},
		and~Khaled~B.~Letaief,~\IEEEmembership{Fellow,~IEEE}% <-this % stops a space
		\thanks{
			Manuscript received October 28, 2016; revised February 21, 2017; accepted
			March 5, 2017. Date of publication XXXXX XX, 2017; date of current
			version XXXXX XX, 2017. This work was supported by the Hong Kong Research Grants Council under Grant No. 16210216 and the U.S.~NSF (grant CCF 1525904).
			This work was presented in part at IEEE Wireless Communications and Networking Conference, Doha, Qatar, Apr. 2016 \cite{mine}.}% <-this % stops a space
		\thanks{X. Yu and J. Zhang with the Department of Electronic and Computer Engineering, the Hong Kong University of Science and Technology (HKUST), Kowloon, Hong Kong (e-mail: \{xyuam, eejzhang\}@ust.hk).
					
			M. Haenggi is with the Department of Electrical Engineering, University of Notre Dame, Notre Dame, IN 46556 USA (e-mail: mhaenggi@nd.edu).
			
			K. B. Letaief is with the Department of Electronic and Computer Engineering, the Hong Kong University of Science and Technology (HKUST), Kowloon, Hong Kong, and also with Hamad bin Khalifa University, Doha, Qatar (e-mail: eekhaled@ust.hk; kletaief@hbku.edu.qa).
			
			Color versions of one or more of the figures in this paper are available online at http://ieeexplore.ieee.org.
			
			Digital Object Identifier 10.1109/JSAC.2017.XXXXXXX
		}
	}% <-this % stops a space
	
	\maketitle
	
	\begin{abstract}
		Millimeter wave (mm-wave) communications is considered a promising technology for 5G networks. Exploiting beamforming gains with large-scale antenna arrays to combat the increased path loss at mm-wave bands is one of its defining features. However, previous works on mm-wave network analysis usually adopted oversimplified antenna patterns for tractability, which can lead to significant deviation from the performance with actual antenna patterns.
		In this paper, using tools from stochastic geometry, we carry out a comprehensive investigation on the impact of directional antenna arrays in mm-wave networks. We first present a general and tractable framework for coverage analysis with arbitrary distributions for interference power and arbitrary antenna patterns. It is then applied to mm-wave ad hoc and cellular networks, where two sophisticated antenna patterns with desirable accuracy and analytical tractability are proposed to approximate the actual antenna pattern. Compared with previous works, the proposed approximate antenna patterns help to obtain more insights on the role of directional antenna arrays in mm-wave networks. In particular, it is shown that the coverage probabilities of both types of networks increase as a non-decreasing concave function with the antenna array size. The analytical results are verified to be effective and reliable through simulations, and numerical results also show that large-scale antenna arrays are required for satisfactory coverage in mm-wave networks.
	\end{abstract}
%	\newpage
	
	\IEEEpeerreviewmaketitle
	
	\section{Introduction}
	\IEEEPARstart{T}{o} meet the ever-increasing demands for high-data-rate multimedia access, the capacity of next-generation wireless networks has to increase exponentially. One promising way to boost the capacity is to exploit new spectrum bands. Recently, millimeter wave (mm-wave) bands from 28 GHz to 300 GHz have been proposed as a promising candidate for new spectrum in 5G networks, which previously were only considered for indoor and fixed outdoor scenarios \cite{hur2013millimeter}. This proposal is supported by recent experiments in the United States and Korea \cite{rappaport2014millimeter,roh2014millimeter}, showing that mm-wave signals can cover up to 200 meters.
	
	Lately, channel measurements have confirmed some unique propagation characteristics of mm-wave signals \cite{rappaportchannel}. It turns out that mm-wave signals are sensitive to blockages, which causes totally different path loss laws for line-of-sight (LOS) and non-line-of-sight (NLOS) mm-wave signals. Furthermore, diffraction and scattering effects are shown to be limited for mm-wave signals. This makes the conventional channel model for sub-6 GHz systems no longer suitable, and thus more sophisticated channel models are needed for the performance analysis of mm-wave networks.
	
	Another distinguishing characteristic of mm-wave signals is the directional transmission. Thanks to the small wavelength of mm-wave signals, large-scale directional antenna arrays can be leveraged to provide substantial array gains and synthesize highly directional beams, which help to compensate for the additional free space path loss caused by the ten-fold increase of the carrier frequency \cite{rappaport2014millimeter}.
	More importantly, different from the rich diffraction and scattering environment in sub-6 GHz systems, directional antennas will dramatically change the signal power, as well as the interference power. In mm-wave networks, the signal or interference power is highly directional and closely related to the angles of departure/arrival (AoDs/AoAs). In particular, the directional antenna array will provide variable power gains corresponding to different AoDs/AoAs. Even a slight shift of AoD/AoA may lead to a large array gain variation. Therefore, it is necessary and critical to incorporate the directional antenna arrays when analyzing mm-wave networks.
	
	%While there have been many studies on coverage analysis for wireless networks in recent years \cite{6042301,chang,haenggi2012stochastic,6928420}, the aforementioned unique properties of mm-wave networks have drawn renewed attention from both academia and industry. Our investigation will focus on the impact of directional antenna arrays, in both mm-wave ad hoc and cellular networks, while taking mm-wave signal propagation characteristics into account.
	
	\subsection{Related Works and Motivation}
	There exist several studies of the coverage performance of mm-wave networks \cite{7094802,venugopal2015interference,robertadhoc,robertcoverage,7105406,7279196,7397837,actualpattern,7370940,7357653,7154396,6824746}.
	Analytical results for coverage and rate coverage probabilities in noise-limited mm-wave networks were presented in \cite{7094802}. Although directional transmission does, to some extent, suppress co-channel interference, a dense deployment is usually required to overcome the blockage in mm-wave networks, which makes mm-wave networks prone to be interference-limited. Hence, only including noise into the coverage analysis is not enough.
	Analytical results on signal-to-interference-plus-noise ratio (SINR) and rate coverage based on a simplified directional antenna pattern were obtained for device-to-device (D2D) \cite{venugopal2015interference}, ad hoc \cite{robertadhoc}, and cellular \cite{robertcoverage,7105406} networks, respectively.
	%Signal propagation in these works was simplified as an omni-directional one and thus did not reflect the propagation characteristics of mm-wave signals.
	To maintain analytical tractability, the antenna pattern was simplified as a \emph{flat-top} pattern, which is a widely used simplification. Since the directional antenna array is a differentiating feature in mm-wave systems, it is crucial and intriguing to accurately incorporate it into the mm-wave network performance analysis.
	
	Basically, the flat-top antenna pattern quantizes the continuously varying antenna array gains in a binary manner. Although it significantly simplifies the analytical derivation, the oversimplified flat-top pattern will lead to pessimistic coverage results, as will be revealed in this paper. Moreover, it is difficult to analyze the impact of directional antenna arrays with the flat-top antenna pattern, as only a few parameters are extracted to abstractly depict the actual antenna pattern. In practice, some critical parameters of the antenna beam pattern such as beamwidth, the $n$-th minor lobe maxima, nulls, and front-back ratio are all determined by the array size. Nevertheless, with the flat-top antenna pattern, these parameters can only be determined qualitatively and inaccurately according to the array size.  As a side effect, the quantized antenna array gain also hinders further investigations of directional antenna arrays. For example, it is difficult to analyze beam misalignment, which is a critical problem in mm-wave networks \cite{hur2013millimeter}.
	
	Recently, there have been some works considering the actual antenna pattern. 
	%An approximative analysis was carried out in \cite{7036065}, which used the maximum antenna array gain for all the interferers to obtain a performance lower bound.
	Two works considered random beamforming in mm-wave networks and used the actual antenna pattern, but only focused on the single link analysis without interference \cite{7279196,7397837}, and adopted some asymptotic approximation in the analysis.
	The actual antenna pattern was adopted in \cite{actualpattern} for evaluating the capacity of an interfered communication link. However, all the interferers were assumed to use the same array gain, which weakens the practicality of the analytical result. An SINR coverage analysis incorporating the actual antenna pattern was carried out in \cite{7370940}. While the coverage probability is analytically given, the multiple integrals (4 nested integrals in the expression) prevent practical evaluation. Also, a Rayleigh fading channel model is not realistic for mm-wave networks due to their poor scattering property. All these works demonstrated that the actual antenna pattern suffers from poor analytical tractability. In addition, there are some works proposing different approximate antenna patterns. In \cite{7357653}, a Gaussian antenna pattern was numerically shown to be a good candidate to approximate the actual antenna pattern but does not lend itself to further analysis. Moreover, though the aforementioned works presented some analytical results with the actual antenna pattern, none of them unraveled how the array size will influence mm-wave networks, which is a critical and unique problem in mm-wave systems, and has only been reported through some simulation works \cite{7154396,6824746}.
	To this end, an antenna pattern that not only approximates the actual antenna pattern accurately and realistically, but also with acceptable analytical tractability, is required to reveal more insights on directional antenna arrays in mm-wave networks.
	
	In summary, there is so far no comprehensive investigation on the impact of directional antenna arrays in mm-wave networks. In this work, we will fill this gap with new analytical results of coverage probabilities that adopt more accurate approximations for the actual antenna pattern.

	\subsection{Contributions}
	We investigate the coverage\footnote{The terminology ``coverage'' is used for both cellular and ad hoc networks.} probabilities in mm-wave networks with a random spatial network model, where transmitters are modeled as a homogeneous Poisson point process (PPP) \cite{haenggi2012stochastic}, and the blockage effect is reflected by a \emph{LOS ball} blockage model \cite{robertcoverage}.
	All the transmitters are assumed to utilize analog beamforming to serve the corresponding users. The main contributions of this work are summarized as follows.
	\begin{itemize}
		\item We first present a general framework for the coverage analysis in mm-wave networks, with arbitrary interference power distributions and antenna patterns, under the assumption that the information signal power is gamma distributed. Compared to previous results, the new expression of the coverage probability is more compact and can be evaluated more efficiently.
		\item Based on the general framework, analytical expressions of the coverage probabilities for both mm-wave ad hoc and cellular networks are provided. For these two types of networks, two approximate antenna patterns are proposed to achieve a good balance between accuracy and analytical tractability. While the proposed approximate antenna patterns are more complicated than the flat-top pattern, our analytical results are more tractable for practical evaluation, thanks to a new approach to deal with gamma distributed signal powers and interferers located in a finite region.
		\item With the highly tractable coverage probabilities at hand, the impact of directional antenna arrays in both mm-wave ad hoc and cellular networks is investigated. We show that the coverage probabilities are monotone increasing functions of the array size. Moreover, the increasing functions are similar in both kinds of networks, which is the product of an exponential and a polynomial function of the inverse of array size. Asymptotic outage probabilities are also derived when the number of antennas goes to infinity, which shows that the asymptotic outage probability is inversely proportional to the array size. This is the first analytical result on the impact of antenna arrays that has been derived in mm-wave networks.
		\item All the analytical results are shown to be computationally efficient through numerical evaluations. Numerical results also show that NLOS signals and NLOS interference have negligible impact on the coverage probability in mm-wave networks. Moreover, the interference power in mm-wave networks is shown to be dominated by the directional antenna array gains, and large-scale directional antenna arrays are needed in mm-wave networks to maintain an acceptable coverage probability. With the increasing network density, the coverage probability has a peak value in mm-wave cellular networks, while it monotonically decreases in ad hoc networks.
	\end{itemize}
	
	\subsection{Organization}
	The remainder of this paper is organized as follows. We shall present the system model in Section \ref{sec_sys}, and a general coverage analysis framework for mm-wave networks is introduced in Section \ref{sec_frame}. Then the coverage probabilities, as well as the impact of directional antenna arrays, for mm-wave ad hoc and cellular networks are derived in Sections \ref{sec_ad} and \ref{sec_cellular}, respectively. Numerical results will be presented in Section \ref{numer}, and conclusions will be drawn in Section \ref{conclu}.
	
	\section{System Model}\label{sec_sys}
	\subsection{Network and Channel Models}\label{II-A}
	We consider downlink transmission in both mm-wave ad hoc and cellular networks. We will first present the common features for both types of networks, and the difference will be specified later. The transmitters are assumed to be distributed according to a homogeneous PPP \cite{haenggi2012stochastic}, which has been shown to be a network model with both reasonable accuracy and analytical tractability \cite{6620915}. As depicted in Fig. 1, we consider the receiver at the origin, which, under an expectation over the point process, becomes the typical receiver. 
	We assume that each receiver has a single receive antenna and is receiving signals from the corresponding transmitter equipped with a directional antenna array composed of $N_\mathrm{t}$ elements. All transmitters operate at a constant power $P_\mathrm{t}$. 
	
\begin{figure}[tbp]
		\centering
		\subfigure
		{
			\centering\includegraphics[height=5.6cm]{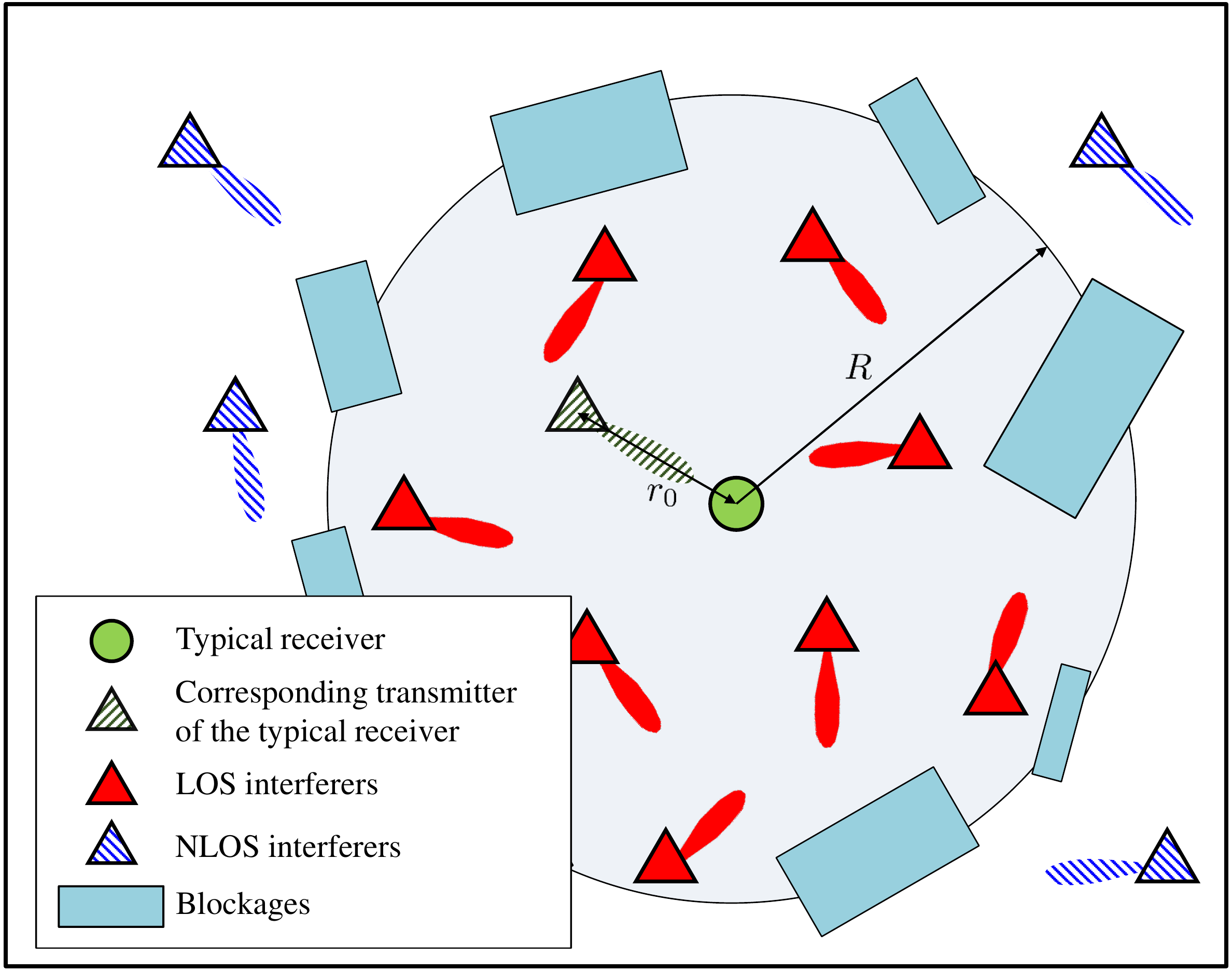}\label{systemmodel}
		}\quad\quad
		\subfigure
		{
			\centering\includegraphics[height=3.6cm]{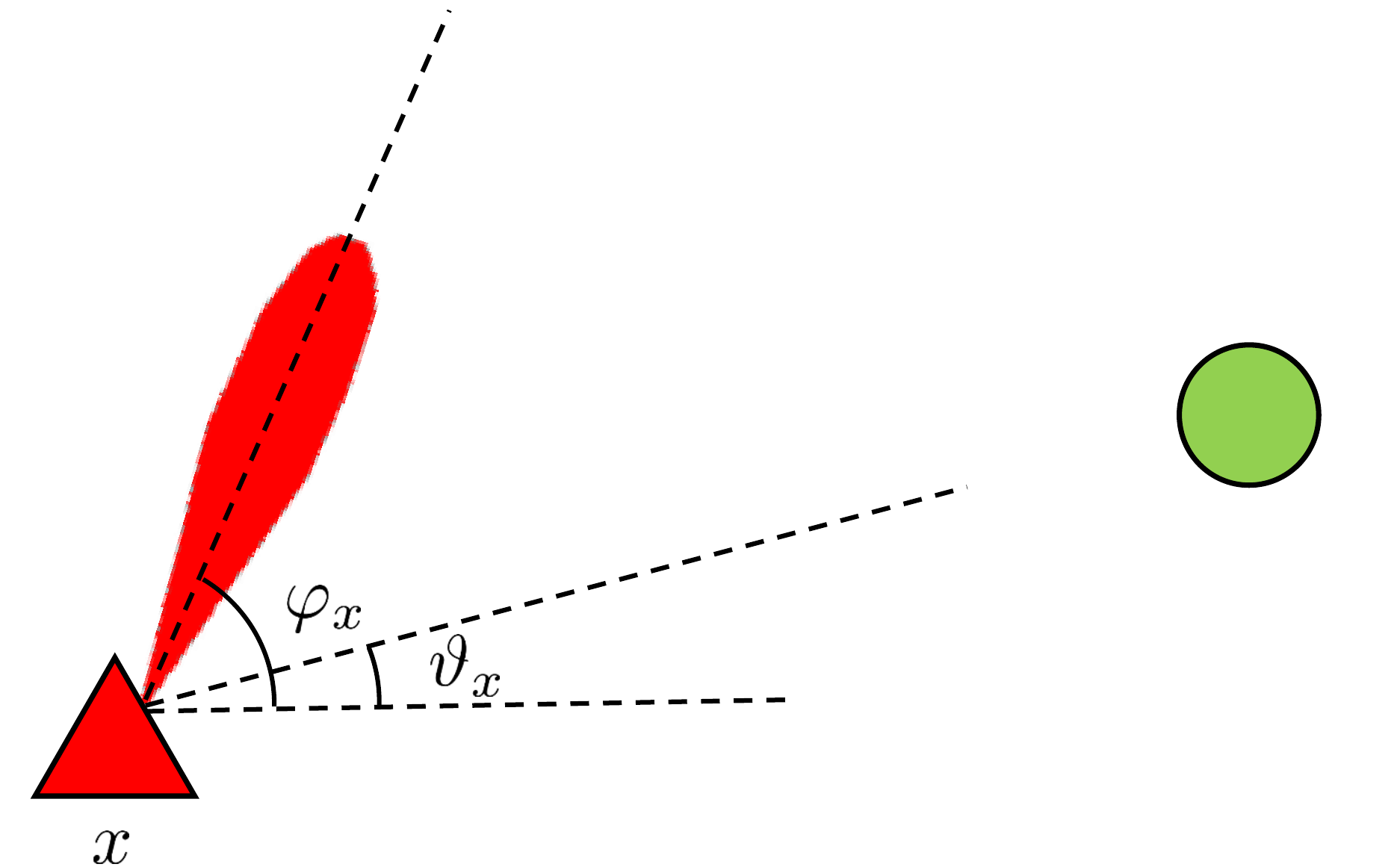}\label{system2}
		}
		\caption{(a): A sample mm-wave network where transmitters are modeled as a PPP. The LOS ball is used to model the blockage effect in the network. (b): Illustration of the spatial AoDs $\vartheta_x$ and $\varphi_x$.}
	\end{figure}

	We use the LOS ball \cite{robertcoverage,7061455} to model the blockage effect as shown in Fig. \ref{systemmodel}. Specifically, we define a LOS radius $R$, which represents the distance between a receiver and its nearby blockages, and the LOS probability of a certain link is one within $R$ and zero outside the radius. Compared with other blockage models adopted in the performance analysis for mm-wave networks, e.g., the 3GPP-like urban micro-cellular model, the LOS ball model has a better fit with real-world blockage scenarios \cite{7593259}. The incorporation of the blockages induces different path loss laws for LOS and NLOS links. 
	It has been pointed out in \cite{robertadhoc,robertcoverage} that NLOS signals and NLOS interference are negligible in mm-wave networks. Hence, we will focus on the analysis where the typical receiver is associated with a LOS transmitter and the interference stems from LOS interferers. The relevant transmitters thus form a PPP, denoted as $\Phi$, with density $\lambda_\mathrm{b}$ in a disk of radius $R$ centered at the origin. In Section \ref{numer}, we will justify the LOS assumption through simulations.
	
	Directional antenna arrays are leveraged to provide significant beamforming gains to overcome the path loss and to synthesize highly directional beams. Universal frequency reuse is assumed, and thus the received signal for the typical receiver is given by
	\begin{equation}
	\begin{split}
		y&=\sqrt{\beta}r_0^{-\frac{\alpha}{2}}\mathbf{h}_{x_0}\mathbf{w}_{x_0}\sqrt{P_\mathrm{t}}s_{x_0}\\
		&\relphantom{=}+\sum_{x\in\Phi^\prime}\sqrt{\beta}\|x\|^{-\frac{\alpha}{2}}\mathbf{h}_{x}\mathbf{w}_x\sqrt{P_\mathrm{t}}s_x+n_0,\label{receivedsignal}
	\end{split}
	\end{equation}
	where $r_0=\|x_0\|$ is the distance between the typical receiver and its corresponding transmitter, while $\|x\|$ is the distance between the transmitter at location $x$ and the typical receiver. The locations of the interfering transmitters are denoted as $\Phi^\prime$, and the channel vector between the interferer and the typical receiver is denoted as $\mathbf{h}_{x}$. The path loss exponent and intercept are symbolized by $\alpha$ and $\beta$ \cite{rappaportchannel}. In addition, the beamforming vector of the transmitter at location $x$ is denoted as $\mathbf{w}_x$, and $n_0$ stands for the additive white Gaussian noise (AWGN) with power being $\sigma^2$.
	
	One main difference between the models for mm-wave ad hoc and cellular networks is the distance $r_0$ between the typical receiver and its corresponding transmitter. In ad hoc networks, each transmitter is assumed to have a corresponding receiver at a fixed distance $r_0$ that is called the \emph{dipole distance}. On the other hand, in cellular networks, the distance $r_0$ between the typical user and its serving base station (BS) is random, due to the random locations of BS and users. We assume that the typical user is associated with its nearest BS, which is commonly adopted in cellular network analysis. The difference in $r_0$ also gives rise to another difference between these two kinds of networks, i.e., the set of interferers $\Phi^\prime$. In ad hoc networks, a dipolar pair is added with the receiver at the origin, and this pair becomes the typical pair. Therefore, $\Phi^\prime=\Phi$. On the other hand, because each user in cellular networks is associated with the nearest BS, which is part of the PPP $\Phi$, the set of interfering BSs $\Phi^\prime=\Phi\backslash\{x_0\}$ forms a PPP conditional on $x_0$ within a ring with inner diameter $r_0$ and outer diameter $R$. 
	
	Next we will present the channel model. Due to high free-space path loss, the mm-wave propagation environment is well characterized by a clustered channel model, i.e., the Saleh-Valenzuela model \cite{rappaportchannel},
	\begin{equation}    
		\mathbf{h}_x=\sqrt{N_\mathrm{t}}\sum_{l=1}^L\rho_{xl}\mathbf{a}_\mathrm{t}^H(\vartheta_{xl}),
	\end{equation}
	where $(\cdot)^H$ symbolizes the conjugate transpose and $L$ is the number of clusters. The complex small-scale fading gain of the $l$-th cluster is denoted as $\rho_{xl}$. Due to the poor scattering environment, especially for LOS signals and interference, the Rayleigh fading assumption commonly used in sub-6 GHz systems no longer holds, which has also been noted in recent works \cite{7593259}. In this paper, we assume, as in \cite{robertcoverage}, that $|\rho_{xl}|$ follows independent Nakagami-$M$ fading for each link.
	
	For mm-wave channels containing LOS components, the effect of NLOS signals is negligible since the channel gains of NLOS paths are typically 20 dB weaker than those of LOS signals \cite{rappaportchannel}. Hence, for the remainder of this paper, we will focus on LOS paths, i.e., $L=1$, and adopt a uniformly random single path (UR-SP) channel model that is commonly used in mm-wave network analysis \cite{7279196,7397837,7160780,6484896}.
	In addition, $\mathbf{a}_t(\vartheta_x)$ represents the transmit array response vector corresponding to the spatial AoD $\vartheta_x$, and it has been shown in \cite[Fig. 3]{mine} that uniform distribution is an excellent approximation for the distribution of spatial AoDs. We consider the uniform linear array (ULA) with $N_\mathrm{t}$ antenna elements. Therefore, the array response vectors are written as
	\begin{equation}
		\begin{split}
			\mathbf{a}_\mathrm{t}(\vartheta_x)=\frac{1}{\sqrt{N_\mathrm{t}}}\left[1,\cdots,e^{{j2\pi k\vartheta_x}},
			\cdots,e^{{j2\pi\left(N_\mathrm{t}-1\right) \vartheta_x}}\right]^T,
		\end{split}
	\end{equation}
	where $\vartheta_x=\frac{d}{\lambda}\cos\phi_x$ is assumed uniformly distributed over $\left[-\frac{d}{\lambda},\frac{d}{\lambda}\right]$, and $0\le k<N_\mathrm{t}$ is the antenna index. Furthermore, $d$, $\lambda$, and $\phi_x$ are the antenna spacing, wavelength, and physical AoD. In order to enhance the directionality of the beam, the antenna spacing $d$ should be no larger than half-wavelength to avoid grating lobes \cite{balanis2005antenna}.
	
	\subsection{Analog Beamforming and Antenna Pattern}\label{II-B}
	While various space-time processing techniques can be applied at each multi-antenna mm-wave transmitter, we focus on analog beamforming, where the beam direction is controlled via phase shifters. Due to the low cost and low power consumption, analog beamforming has already been adopted in commercial mm-wave systems such as WiGig (IEEE 802.11ad) \cite{rappaport2014millimeter}. Assuming the spatial AoD of the channel between the transmitter at location $x$ and its serving user is $\varphi_x$, the optimal analog beamforming vector is well known and given by
	\begin{equation}
		\mathbf{w}_x=\mathbf{a}_\mathrm{t}(\varphi_x),\label{beamalign}
	\end{equation}
	which means the transmitter should align the beam direction exactly with the AoD of the channel to obtain the maximum power gain.
	
\begin{figure*}
		\centering
		\subfigure[Visualization of four different antenna patterns when $N_\mathrm{t}=64$.]
		{
			\centering\includegraphics[height=5.6cm]{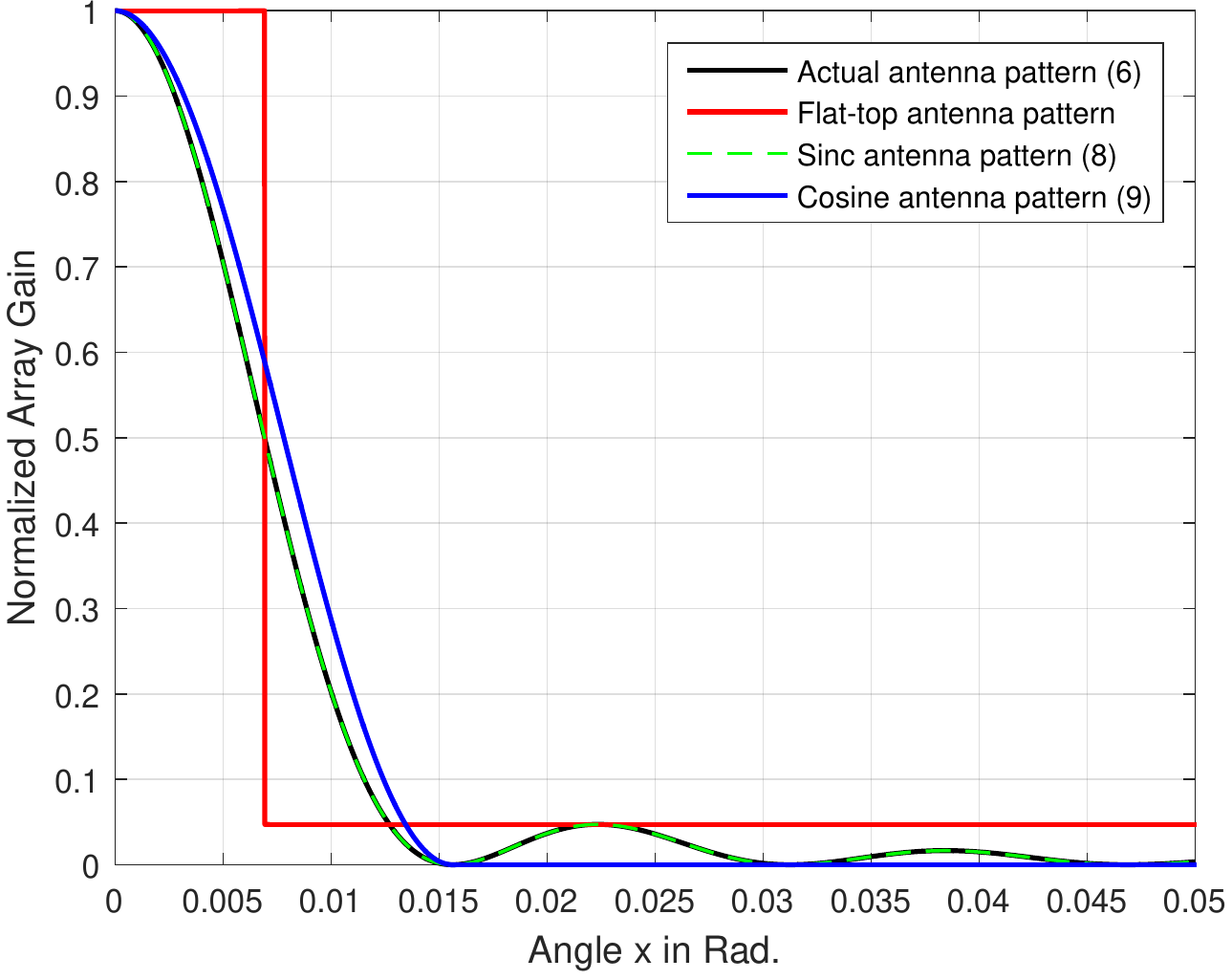}\label{motivation1}
		}
		\subfigure[Coverage probability evaluations using four different antenna patterns in mm-wave cellular networks when $R=200$ m, $N_\mathrm{t}=64$, $\lambda_\mathrm{b}=1\times 10^{-3}$ m$^{-2}$, $M=3$, and $\alpha=2.1$.]
		{
			\centering\includegraphics[height=5.6cm]{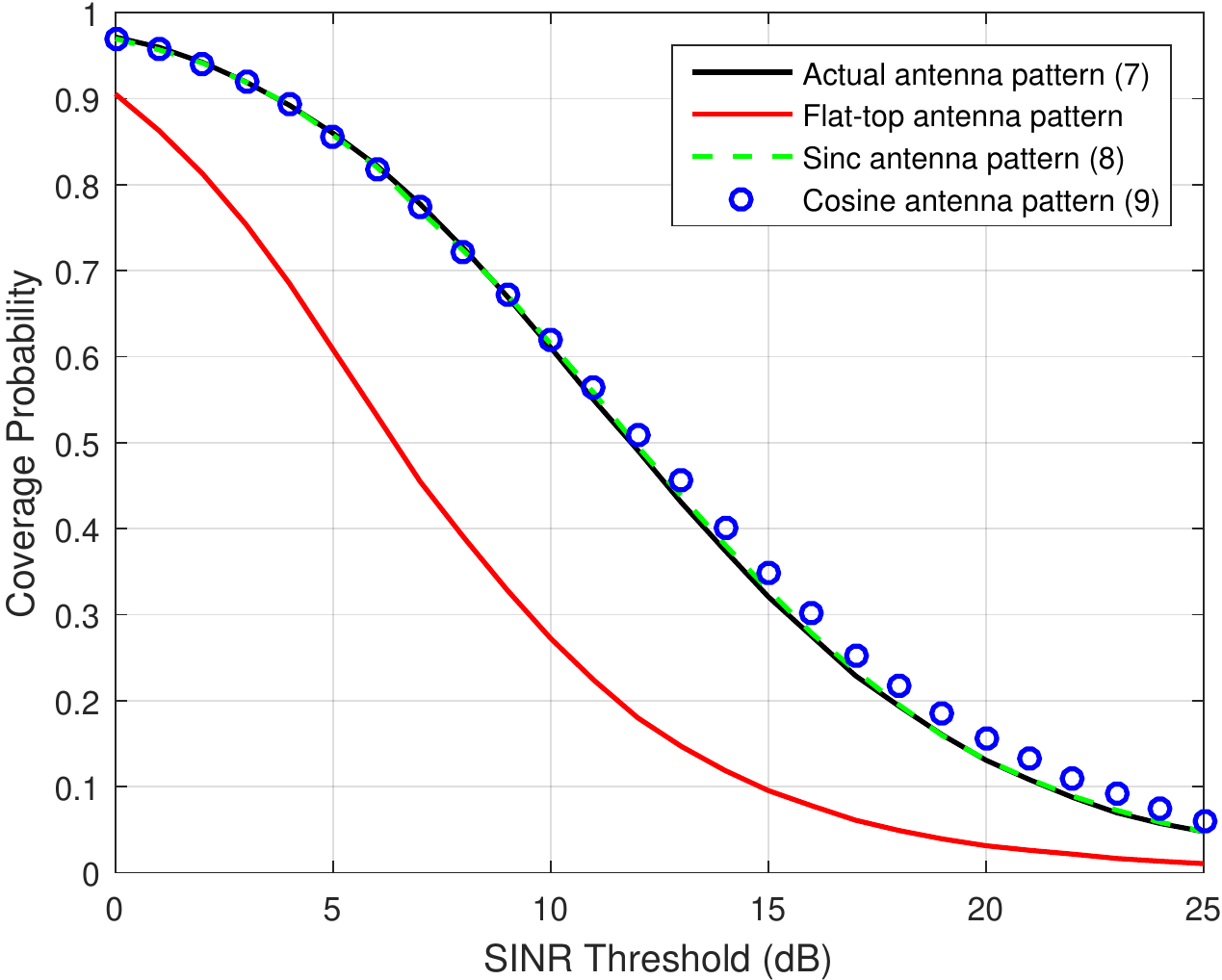}\label{motivation2}
		}
		\caption{The comparisons between different approximate antenna patterns.}\label{motivation}
	\end{figure*}

	As shown in Fig. \ref{system2}, based on the optimal analog beamforming vector \eqref{beamalign}, for the typical receiver, the product of small-scale fading gain and beamforming gain of the transmitter at location $x$ is given by
	\begin{equation}
		\left|\mathbf{h}_x\mathbf{w}_x\right|^2=N_\mathrm{t}\left|\rho_x\right|^2\left|\mathbf{a}_\mathrm{t}^H(\vartheta_x)\mathbf{a}_\mathrm{t}(\varphi_x)\right|^2,
		% \triangleq N_\mathrm{t}\left|\rho_x\right|^2G_\mathrm{act}(\vartheta_x-\varphi_x),
	\end{equation}
	where $\left|\rho_x\right|^2$ is the power gain of small-scale fading. By defining the array gain function $G_\mathrm{act}(x)$ as
	\begin{equation}
		G_\mathrm{act}(x)\triangleq\frac{\sin^2\left(\pi N_\mathrm{t}x\right)}{N_\mathrm{t}^2\sin^2\left(\pi x\right)},
	\end{equation} 
	the normalized array gain of the transmitter at location $x$ can be expressed as
	\begin{equation}
	\begin{split}
		\left|\mathbf{a}_\mathrm{t}^H(\vartheta_x)\mathbf{a}_\mathrm{t}(\varphi_x)\right|^2&=
		\frac{1}{N_\mathrm{t}^2}\left|\sum_{i=0}^{N_\mathrm{t}-1}{e^{j2\pi i(\vartheta_x-\varphi_x)}}\right|^2\\
		&=\frac{\sin^2\left[\pi N_\mathrm{t}(\vartheta_x-\varphi_x)\right]}{N_\mathrm{t}^2\sin^2\left[\pi (\vartheta_x-\varphi_x)\right]}= G_\mathrm{act}(\vartheta_x-\varphi_x),
		\label{sinsin}
	\end{split}
	\end{equation}
	where $\vartheta_x$ and $\varphi_x$ are independent uniformly distributed random variables over $\left[-\frac{d}{\lambda},\frac{d}{\lambda}\right]$. The array gain function in \eqref{sinsin} is a normalized \emph{Fej\'er kernel} with factor $\frac{1}{N_\mathrm{t}}$ and is referred as the \emph{actual antenna pattern}.
	In fact, the distribution of $\vartheta_x-\varphi_x$ in \eqref{sinsin} is uniform, which is stated in the following lemma. Note that this substitution will not change the overall distribution of the array gain.
	\begin{lemma}
		The array gain $G_\mathrm{act}(\vartheta_x-\varphi_x)$ is equal in distribution to $G_\mathrm{act}\left(\frac{d}{\lambda}\theta_x\right)$, where $\theta_x$ is a uniformly distributed random variable over $[-1,1]$.
	\end{lemma}
	\begin{IEEEproof}
		The proof is based on the uniform distribution of $\vartheta_x$ and $\varphi_x$, and the periodic property of the function $e^{j2\pi x}$ in \eqref{sinsin}. The proof has been established in \cite[Appendix A]{7279196}.
	\end{IEEEproof}
	Although the Fej\'er kernel has a relatively simple analytical form, it does not lend itself to further analysis due to the sine functions in both the numerator and denominator, which calls for an approximate antenna pattern with both accuracy and tractability in performance analysis of mm-wave networks.
	Next we will introduce two new approximate antenna patterns, as well as the flat-top antenna pattern, which has been widely used in existing works. Fig. \ref{motivation} visualizes these antenna patterns and evaluates the coverage probabilities with different antenna patterns through simulation.
	
	\emph{1) Flat-top antenna pattern}:
	Most of the existing works \cite{venugopal2015interference,robertadhoc,robertcoverage,7105406} adopt this simplified antenna pattern in the coverage analysis, where the array gains within the half-power beamwidth (HPBW) \cite{balanis2005antenna} are assumed to be the maximum power gain, and the array gains corresponding to the remaining AoDs are approximated to be the first minor maximum gain of the actual antenna pattern. While this simple approximation is highly tractable, it introduces huge discrepancies when we evaluate the network coverage probability, as shown in Fig. \ref{motivation2}\footnote{The gap can be narrowed by  heuristically choosing different parameters for the flat-top pattern, e.g., beamwidth and front-back ratio, but the overall shape of the coverage probability remains different.}. 
	
	\emph{2) Sinc antenna pattern}: Instead of the actual antenna pattern, a tight lower bound is widely adopted for the numerical analysis in antenna theory. Since the antenna spacing $d$ is usually no larger than half-wavelength to avoid grating lobes, and $\sin x\simeq x$ for small $x$, the array gain function can be approximately expressed as \cite[Equation (6-10d)]{balanis2005antenna}
	\begin{equation}
		G_\mathrm{sinc}(x)
		\triangleq\frac{\sin^2\left(\pi N_\mathrm{t}x\right)}{\left(\pi N_\mathrm{t} x\right)^2},
	\end{equation}
	which is a squared sinc function. The accuracy of this tight lower bound is shown in \cite[Appendix I,II]{balanis2005antenna}. In Fig. \ref{motivation1}, it turns out that the sinc antenna pattern is almost the same as the actual antenna pattern, and there is almost no error when using this approximate antenna pattern to investigate the coverage probability, as illustrated in Fig. \ref{motivation2}. Moreover, note that the sinc function is more tractable due to the absence of the sine function in the denominator, compared to the actual antenna pattern.
	
	\emph{3) Cosine antenna pattern}: 
	Another antenna pattern approximation is based on the cosine function as follows
	\begin{equation}\label{approxpattern}
		G_\mathrm{cos}(x)=
		\begin{cases}
			\cos^2\left(\frac{\pi N_\mathrm{t}}{2}x\right)&|x|\le\frac{1}{N_\mathrm{t}},\\
			0&\text{otherwise},
		\end{cases}
	\end{equation}
	where the nonzero part is an elementary function with better analytical tractability. In Fig. \ref{motivation1}, we observe that the cosine antenna pattern provides a good approximation for the main lobe gains while sacrificing the accuracy for the side lobe ones. When incorporated into the coverage probability, the cosine antenna pattern has negligible gap between the actual antenna pattern, which can be viewed as a desirable trade-off between accuracy and tractability in performance analysis for mm-wave networks.
	
	Fig. \ref{motivation} shows that the sinc and cosine antenna patterns are more accurate. In particular, they are superior to the flat-top pattern since they accurately capture the impact of directional antenna arrays in mm-wave networks. In particular, given the operating frequency and the antenna spacing, the antenna pattern is critically determined by the array size. In the flat-top pattern, however, it is very difficult to quantitatively and accurately depict the variation of the HPBW and the first minor maximum for different array sizes and AoDs. Moreover, the binary quantization of the array gains cannot reflect the roll-off characteristic of the actual antenna pattern and therefore is unable to provide different array gains for various AoDs. In other words, the flat-top antenna pattern obliterates the possibility of analyzing the impact of directional antenna arrays, which is a critical and unique issue in mm-wave systems. On the contrary, the sinc and cosine antenna patterns are explicit functions of the array size, which makes it possible to investigate the relation between the coverage probability and the directional antenna arrays. The sinc and cosine antenna patterns will be adopted in the coverage analysis for mm-wave ad hoc and cellular networks in Sections \ref{sec_ad} and \ref{sec_cellular}, respectively.
	
	\section{A General Framework for Coverage Analysis of mm-wave Networks}\label{sec_frame}
	In this section, we will develop a general framework for the coverage analysis of mm-wave networks. The main result is a tractable expression for the coverage probability, for arbitrary antenna patterns and interference distributions. It will then be applied in the following two sections to evaluate mm-wave ad hoc networks and mm-wave cellular networks.
	\subsection{Signal-to-interference-plus-noise Ratio (SINR) Analysis}\label{III-A}
	We assume that each transmitter has full information about the AoD of the channel between itself and its serving user, which can be obtained through sophisticated beam training protocols \cite{rappaport2014millimeter}. The transmitters can align the beam to the AoD direction, according to \eqref{beamalign}, using analog beamforming to obtain the maximum antenna array gain. The SINR at the typical receiver is then given by
	\begin{equation}
		\mathrm{SINR}=\frac{P_\mathrm{t}N_\mathrm{t}|\rho_{x_0}|^2\beta r_0^{-\alpha}}{\sigma^2 + \sum_{x\in\Phi^\prime}P_\mathrm{t}g_x\beta \|x\|^{-\alpha}}= \frac{|\rho_{x_0}|^2 r_0^{-\alpha}}{\sigma_\mathrm{n}^2 + \sum_{x\in\Phi^\prime}g_x \|x\|^{-\alpha}},\label{generalSINR}
	\end{equation}
	where $\sigma_\mathrm{n}^2=\frac{\sigma^2}{\beta P_\mathrm{t}N_\mathrm{t}}$ is the normalized noise power and $g_x$ is the channel gain from the interfering transmitter at location $x$, including both the small-scale fading gain and the directional antenna array gain\footnote{Since $g_x$ is an arbitrary channel gain, when normalizing the noise power by $N_\mathrm{t}$, we abbreviate the normalized channel gain $\frac{g_x}{N_\mathrm{t}}$ as $g_x$ with a slight abuse of notation.}. In this section, we assume $(g_x)_{x\in\Phi^\prime}$ is a family of non-negative random variables with independent and identically distributions, which will be specified in the following two sections.
	
	Besides the complicated directional antenna array gains, there is another difficulty when calculating the SINR distribution. Note that the channel gain for the signal $|\rho_{x_0}|^2$ follows a gamma distribution $\mathrm{Gamma}\left(M,\frac{1}{M}\right)$, where $M$ is the Nakagami parameter. Compared with the exponential distributed power gain induced by Rayleigh fading, this gamma distribution brings additional challenges into the derivation. Note that the gamma distribution for the signal power widely appears when evaluating various multi-antenna systems. Many previous works illustrated that the signal power is gamma distributed considering more general transmission techniques, e.g., maximal ratio transmission \cite{chang}, or other network settings such as heterogeneous networks \cite{chang1}.  
	%The other difficulty is the spatial distribution of the concerned interferers, i.e., the PPP $\Phi$ in a finite area with radius $R$. In conventional coverage analysis, the interferers are located in an infinite area, which greatly simplifies the derivation due to the asymptotic property of the decaying exponential function \cite{6042301}. Nevertheless, with a finite boundary, deriving the distribution of the interference would call for extra demanding efforts. We will demonstrate how we solve these two analytical challenges in coverage analysis for mm-wave networks in the next subsection. 
	
	\subsection{Coverage Analysis Framework}\label{III-B}
	The coverage probability, defined as the probability that the received SINR is greater than a certain threshold $\tau$, is written as
	\begin{equation}
	\begin{split}
		p_\mathrm{c}(\tau)&=\mathbb{P}\left(\frac{|\rho_{x_0}|^2 r_0^{-\alpha}}{\sigma_\mathrm{n}^2 + \sum_{x\in\Phi^\prime}g_x \|x\|^{-\alpha}}>\tau\right)\\
		&=\mathbb{P}\left[|\rho_{x_0}|^2>\tau r_0^\alpha\left(\sigma_\mathrm{n}^2 +I\right)\right],\label{coverageprob}
	\end{split}
	\end{equation}
	where $I=\sum_{x\in\Phi^\prime}g_x \|x\|^{-\alpha}$.
	As mentioned before, one main difficulty of the analysis comes from the gamma distributed random variable $|\rho_{x_0}|^2$. In previous works that investigated the coverage analysis for mm-wave networks \cite{mine, venugopal2015interference,robertadhoc,robertcoverage}, an upper bound for the cumulative probability function (cdf) of a normalized gamma random variable was adopted. In contrast, in this paper, we will derive an exact expression for this probability. The coverage probability \eqref{coverageprob} is firstly rewritten as
	\begin{equation}
	\begin{split}
		p_\mathrm{c}(\tau)&\overset{(a)}{=}\mathbb{E}_{r_0}\left\{\sum_{n=0}^{M-1}\frac{(M\tau r_0^\alpha)^n}{n!}\mathbb{E}_I\left[(\sigma_\mathrm{n}^2+I)^ne^{-M\tau r_0^\alpha (\sigma_\mathrm{n}^2+I)}\right]\right\}\\
		&=\mathbb{E}_{r_0}\left[\sum_{n=0}^{M-1}\frac{(-s)^n}{n!}\mathcal{L}^{(n)}(s)\right],
		\label{nthde}
	\end{split}
	\end{equation}
	where $s=M\tau r_0^\alpha$, $\mathcal{L}(s)=e^{-s\sigma_\mathrm{n}^2}\mathbb{E}_I\left[e^{-sI}\right]$ is the Laplace transform of noise and interference. The variable $r_0$ is random in cellular networks but deterministic in ad hoc ones. The notation $\mathcal{L}^{(n)}(s)=(-1)^n\mathbb{E}_I\left[(\sigma_\mathrm{n}^2+I)^ne^{-s(\sigma_\mathrm{n}^2+I)}\right]$ stands for the $n$-th derivative of $\mathcal{L}(s)$, and step $(a)$ is from the cdf of a gamma random variable.
	
	Next, we will derive the coverage probability based on the expression \eqref{nthde}. In particular, we will show that, for arbitrary distributions of the channel gain, the $n$-th derivative of the Laplace transform can be expressed in a recursive form. Afterwards, the coverage probability can be expressed by the induced $\ell_1$-norm of a lower triangular Toeplitz matrix. This approach yields a more compact analytical result for the coverage probability than previous works, thanks to the more delicate handling of the gamma distributed fading gain. More importantly, this framework enables us to perform further analyses of mm-wave networks, e.g., to investigate the impact of directional antenna arrays in later sections, which cannot be unraveled from existing works. 
	
	The first step is to derive the Laplace transform $\mathcal{L}(s)$, given in the following lemma. As mentioned in Section \ref{II-A}, we focus on the LOS interference within the LOS radius $R$ in the following derivation.
	\begin{lemma}Assuming a lower bound $\kappa$ on the distance between the typical receiver and the nearest interferer, the Laplace transform of noise and interference is
		\begin{align}
			\label{Ls}\nonumber
			\mathcal{L}(s)&=\exp\bigg(-s\sigma_\mathrm{n}^2-\pi\lambda_\mathrm{b}\Big\{R^2- \kappa^2+\delta\kappa^2\mathbb{E}_{g}\left[\mathrm{E}_{1+\delta}(s\kappa^{-\alpha} g)\right]\\
			&\relphantom{=}-\delta R^2\mathbb{E}_{g}\left[\mathrm{E}_{1+\delta}(sR^{-\alpha}g)\right]\Big\}\bigg)\nonumber\\
			&\triangleq \exp\left\{\eta(s)\right\},
		\end{align}
		where $\delta=\frac{2}{\alpha}$, $\mathrm{E}_p(z)$ is the generalized exponential integral \cite[Page \text{xxxv}]{zwillinger2014table}, and $g$ is the channel gain  that is distributed as all the $g_x$ in \eqref{generalSINR}.\label{lem1}
	\end{lemma}
	\begin{IEEEproof}
		The Laplace transform of the interference $I$, denoted as $\mathcal{L}_I(s)$, is well known and is written as \cite{6042301}
		\begin{equation}\label{eq14}
		\begin{split}
			\mathcal{L}_I(s)&=\mathbb{E}_I\left[e^{-sI}\right]\\
			&=\exp\left\{-2\pi\lambda_\mathrm{b}\int_\kappa^R{\left(1-\mathbb{E}_{g}[\exp(-sgx^{-\alpha})]\right)}x\mathrm{d}x\right\}.
		\end{split}
		\end{equation}
		Note that the expectation over $g$ is another integral, so \eqref{eq14} involves a double integral. Since the integration function is integrable, according to Fubini's theorem, we can swap the order of the expectation and integration, and part of the exponent of $\mathcal{L}_I(s)$ can be recast as
		\begin{align}
			&\relphantom{=}2\int_\kappa^R{\left(1-\mathbb{E}_{g}[\exp(-sgx^{-\alpha})]\right)}x\mathrm{d}x\nonumber\\
			&=2\mathbb{E}_{g}\left[\int_\kappa^R\left[1-\exp(-sgx^{-\alpha})\right]x\mathrm{d}x\right]\label{swap}\\
			\nonumber&= R^2- \kappa^2+\mathbb{E}_{g}\left[(sg)^{\delta}\int_{sgR^{-\alpha}}^{sg\kappa^{-\alpha}} e^{-t}\mathrm{d}t^{-\delta}\right]\\
			&= R^2- \kappa^2+\delta\kappa^2\mathbb{E}_{g}\left[\mathrm{E}_{1+\delta}(s\kappa^{-\alpha}g)\right]-\delta R^2\mathbb{E}_{g}\left[\mathrm{E}_{1+\delta}(sR^{-\alpha}g)\right].\label{intlap}
		\end{align}
		By substituting \eqref{intlap} into \eqref{eq14}, the Laplace transform $\mathcal{L}(s)$ in Lemma \ref{lem1} can be obtained.
	\end{IEEEproof}
	Calculating the Laplace transform in PPP usually involves two expectation operations over the interferers' locations and channel gains, respectively. 
	Note that in the derivation of Lemma \ref{lem1}, we first take the expectation over the interferers' locations and then average over the channel gains as shown in \eqref{swap}, which is in the reverse order compared to the conventional derivation \cite{6042301} and existing works in mm-wave networks \cite{mine, venugopal2015interference,robertadhoc,robertcoverage}. The reason why we perform these two expectations in this order is that, in mm-wave networks, the distribution of the channel gains involving the directional antenna array gains are much more complicated than that in sub-6 GHz networks, and therefore we take it as the latter step to maintain the analytical tractability. In later sections we will see the benefits of this swapping.
	
	Based on the Laplace transform derived in Lemma \ref{lem1}, the coverage probability is given in the following theorem.
	\begin{theorem}The coverage probability \eqref{coverageprob} is given by
		\begin{equation}
			p_\mathrm{c}(\tau)=
			\begin{dcases}
				\left\Vert\exp\left(\mathbf{C}_M\right)\right\Vert_1&\mathrm{ad\,\,hoc},\\
				\int_0^Rf_{r_0}(r)\left\Vert\exp\left\{\mathbf{C}_M(r)\right\}\right\Vert_1\mathrm{d}r&\mathrm{cellular},\label{frameexpr}
			\end{dcases}
		\end{equation}
		where $f_{r_0}(r)$ is the probability density function (pdf) of the distance between the typical receiver and its associated transmitter, and $\mathbf{C}_M$ is an $M\times M$ lower triangular Toeplitz matrix
		%\begin{equation}
		%\mathbf{C}_M=\left[ {\begin{array}{*{20}{c}}
		%c_0&{}&{}&{}&{}\\
		%c_1&c_0&{}&{}&{}\\
		%c_2&c_1&c_0&{}&{}\\
		%\vdots &{}&{}& \ddots &{}\\
		%c_{M-1}&\cdots&c_2& c_1 &c_0
		%\end{array}} \right],\label{topmatrix}
		%\end{equation}
		\begin{equation}
			\mathbf{C}_M=\left[{\begin{IEEEeqnarraybox*}[][c]{,c/c/c/c/c,}
					c_0&{}&{}&{}&{}\\
					c_1&c_0&{}&{}&{}\\
					c_2&c_1&c_0&{}&{}\\
					\vdots &{}&{}& \ddots &{}\\
					c_{M-1}&\cdots&c_2& c_1 &c_0
			\end{IEEEeqnarraybox*}} \right],\label{topmatrix}
		\end{equation}
		whose nonzero entries are determined by
		\begin{equation}
			c_k=\frac{(-s)^k}{k!}\eta^{(k)}(s),
		\end{equation}\label{th1}
		and $c_k>0$ for $k\ge1$.
	\end{theorem}
	\begin{IEEEproof}
		See Appendix \ref{AB}.
	\end{IEEEproof}
	As stated in Section \ref{III-A}, the main assumption in Theorem \ref{th1} is the gamma distributed signal power, and this theorem holds for arbitrary interference distributions and antenna patterns. Furthermore, note that we adopt a general exponent $\eta(s)$ of the Laplace transform $\mathcal{L}(s)$, and thus Theorem \ref{th1} can be viewed as a generalization of the results in \cite{chang,chang1}. When the exponent $\eta(s)$ is specified as \cite[Equation (35)]{chang} and \cite[Equation (36)]{chang1} according to different network settings and fading assumptions, Theorem \ref{th1} specializes to the expressions therein. In mm-wave networks, the channel gain $g$ not only includes the small-scale fading gain, but also the directional antenna array gain. With Theorem \ref{th1} at hand, in order to obtain the specific coverage probability expression for a certain kind of channel gain $g$, the only parameters required to be determined are the entries $\{c_k\}_{k=0}^{M-1}$ in the matrix $\mathbf{C}_M$. While we focus on analog beamforming, the framework proposed in this section is also applicable for mm-wave networks adopting other transmission techniques, e.g., hybrid precoding \cite{el2014spatially,7397861}. In the following two sections, we shall derive the coverage probabilities for different network settings and antenna patterns.
	
	\section{Coverage Analysis for mm-Wave Ad Hoc Networks}\label{sec_ad}
	Millimeter wave communications has been proposed as a promising technique for next-generation ad hoc networks with short-range transmission, e.g., military battlefield networks \cite{5273811}, high-fidelity video transmission \cite{rappaport2014millimeter}, and D2D networks \cite{7010536}. In this section, we will first derive an analytical expression of the coverage probability for mm-wave ad hoc networks, based on which we will then investigate the critical role of directional antenna arrays in such networks.
	\subsection{Coverage Analysis}\label{IV-A}
	In mm-wave ad hoc networks, a dipole model is adopted, where the communication distance between the typical receiver and its associated transmitter is assumed to be fixed as the dipole distance \cite{haenggi2012stochastic}. As mentioned in Section \ref{II-A}, we assume that the typical dipole pair is in the LOS condition, i.e., $r_0\le R$. In fact, if the typical receiver is associated with a NLOS transmitter out of the LOS radius, due to the huge path loss and high noise power at mm-wave bands, the coverage probability will be fairly low (close to zero) for a practical SINR threshold, and therefore with little analytical significance. Furthermore, in ad hoc networks, the nearest interferer can be arbitrarily close to the typical receiver, i.e., $\kappa=0$. According to \eqref{generalSINR}, the received SINR is given by
	\begin{equation}
		\mathrm{SINR}=\frac{|\rho_{x_0}|^2r_0^{-\alpha}}
		{\sigma^2_n + \sum_{x\in\Phi}|\rho_x|^2G_\mathrm{act}\left(\frac{d}{\lambda}\theta_x\right) \|x\|^{-\alpha}}.
	\end{equation}
	As mentioned in Section \ref{II-B}, the sinc antenna pattern is an excellent approximation of the actual antenna pattern with better analytical tractability, so we propose to adopt it in the analysis of mm-wave ad hoc networks.
	
	Note that in Section \ref{III-B}, we have pointed out that the main task to derive the coverage probability $p_\mathrm{c}(\tau)$ is to determine the entries in the matrix $\mathbf{C}_M$. The channel gain $g$ is the product of the gamma distributed small-scale fading gain $|\rho_x|^2$ and the directional antenna array gain $G_\mathrm{sinc}\left(\frac{d}{\lambda}\theta_x\right)$. First, a unique property of the directional array gain with the sinc antenna pattern is presented to help derive the coverage probability.
	\begin{lemma}\label{lem3}
		For $p\in\mathbb{Z}^+$,
		\begin{equation}
			\int_0^\infty \left(\frac{\sin x}{x}\right)^{2p}\mathrm{d}x=\frac{\pi}{2(2p-1)!}{{2p-1}\bangle{p-1}},
		\end{equation}
		where $n\bangle k$ are the Eulerian numbers, i.e., ${n\bangle k}=\sum_{j=0}^{k+1}(-1)^j{{n+1}\choose j}(k-j+1)^n$.
	\end{lemma}
	\begin{IEEEproof}
		The proof can be found in \cite[Lemma 2]{mine}.
	\end{IEEEproof}
	
	Based on Lemma \ref{lem3}, a lower bound of the coverage probability with the sinc antenna pattern is derived in the following proposition.
	\begin{prop}\label{th2}The coverage probability of mm-wave ad hoc networks with the sinc antenna pattern is tightly lower bounded by
		\begin{equation}\label{adcov}
			p_\mathrm{c}^{\mathrm{sinc}}(\tau)\ge\left\Vert\exp\left(\frac{1}{N_\mathrm{t}}\mathbf{C}_M\right)\right\Vert_1.
		\end{equation}
		The coefficients in $\mathbf{C}_M$ are given by
			\begin{equation}
			\begin{split}
			c_k&=\Bigg[\frac{\pi R^2\lambda_\mathrm{b}\lambda}{\alpha d}\sum_{p=\max\{1,k\}}^\infty\frac{(-\tau r_0^\alpha)^p{{2p-1}\bangle{p-1}} \Gamma(M+p)}{R^{\alpha p}(2p-1)!(p-k)!\left(p-\delta\right)\Gamma(M)}\\
			&\relphantom{=}-\frac{\delta\lambda_\mathrm{b}\lambda}{ d}\left(\delta\right)_k\Gamma\left(-\delta\right)\frac{\Gamma\left(M+\delta\right)}{\Gamma(M)}\tau^{\delta}r_0^2\xi\\
			&\relphantom{=}+\mathbf1(k\le1)\frac{\tau Mr_0^\alpha\sigma^2}{\beta P_\mathrm{t}}\Bigg]\times\frac{(-1)^{k+1}}{k!},
			\end{split}
			\end{equation}
		where $\Gamma(\cdot)$ denotes the gamma function, $(x)_n$ represents the falling factorial, $\mathbf{1}(\cdot)$ is the indicator function, and
		\begin{equation}
			\xi=\int_{0}^\infty\left|\frac{\sin x}{x}\right|^{2\delta}\mathrm{d}x.
			%	,\quad U(k)=\begin{cases}
			%	1&k=0,\\
			%	k&k\ge1.
			%	\end{cases}
		\end{equation}
	\end{prop}
	\begin{IEEEproof}
		See Appendix \ref{AD}.
	\end{IEEEproof}
	
	\emph{Remark 1:} According to recent mm-wave channel measurements \cite{rappaportchannel}, the path loss exponent $\alpha$ is less than 3, which ensures the convergence of $\xi$.
	
	\emph{Remark 2:} Although the expressions in Proposition \ref{th2} involve a summation of infinitely many terms, it turns out that, in practical evaluation, the series converges quickly, and the high-order terms contribute little to the sum. Hence, using a finite number of terms is sufficient for numerical computation. In addition, $\xi$ only depends on the path loss exponent $\alpha$ and can easily be evaluated numerically and offline. Overall, the expression in Proposition \ref{th2} is much easier to evaluate than existing results \cite{robertadhoc,venugopal2015interference} that contain multiple nested integrals.
	
	\emph{Remark 3:} Note that the derivation in Appendix \ref{AD} is based on the Laplace transform provided in Lemma \ref{lem1}, where we swap the order of two expectations as mentioned in Section \ref{III-B}. 
	%In fact, an unbounded integral will appear if conventional approaches in sub-6 GHz systems \cite{6042301} and existing works in mm-wave systems \cite{mine, venugopal2015interference,robertadhoc,robertcoverage} are applied. 
	With the help of this swapping operation, we are able to derive a more tractable expression for the Laplace transform, which verifies the benefits and superiority of the proposed analytical framework.
	
	\emph{Remark 4:} For a given coverage probability, the maximum transmitter density can be numerically determined by Proposition 1.
	
	\subsection{Impact of Directional Antenna Arrays}\label{IV-B}
	Next we investigate how directional antenna arrays affect the coverage probability in mm-wave networks. Increasing the array size enhances the signal quality, but may also increase interference power. The overall effect is revealed in the following corollary.
	\begin{corollary}\label{coro1}
		The tight lower bound of the coverage probability \eqref{adcov} is a non-decreasing concave function with the array size, and it can be rewritten as
		\begin{equation}\label{eqcoro1}
			p_\mathrm{c}^{\mathrm{sinc}}(t)\ge e^{c_0t}\left(1+\sum_{n=1}^{M-1}\beta_nt^n\right),
		\end{equation}
		where $t=\frac{1}{N_\mathrm{t}}$, and
		\begin{equation}
			\beta_n=
			\frac{\left\Vert \left(\mathbf{C}_M-c_0\mathbf{I}_M\right)^n\right\Vert_1}{n!}\quad n\ge1.
		\end{equation}
		When $t\to0$, i.e., $N_\mathrm{t}\to\infty$, the asymptotic outage probability is given by
		\begin{equation}\label{asymp}
			\tilde{p}_\mathrm{o}^\mathrm{sinc}(t)\sim\frac{\mu}{N_\mathrm{t}},
		\end{equation}
		where $\mu=-\sum_{n=0}^{M-1}c_n>0$. 
	\end{corollary}
	\begin{IEEEproof}
		See Appendix \ref{AE}.
	\end{IEEEproof}
	It can be seen in Corollary \ref{coro1} that $p_\mathrm{c}^\mathrm{sinc}(t)\to 1$ as $t\to0$ for all network parameters. Hence, for any desired coverage requirement $1-\epsilon$, there exists a minimum antenna array size $N_\mathrm{t}$ that can satisfy it regardless of the other network parameters, 
	which can be numerically determined by Corollary \ref{coro1}.
	The lower bound in Corollary \ref{coro1} indicates how antenna arrays affect the coverage probability. From Corollary \ref{coro1}, we discover that increasing the directional antenna array size will definitely benefit the coverage probability in ad hoc networks. Later in Section \ref{numer} we show that the result is tight
	through simulations. Moreover, we see that the lower bound is a product of an exponential function and a polynomial function of order $M-1$ of the inverse of the array size $t$. For the special case that $M=1$, i.e., Rayleigh fading channel, the lower bound reduces to an exponential one. 
	The asymptotic coverage probability \eqref{asymp} shows that the asymptotic outage probability is inversely proportional to the array size.
	To the best of the authors' knowledge, this is the first analytical result on the impact of antenna arrays in mm-wave network analysis. 
	%Furthermore, with the monotonicity property in Corollary \ref{coro1}, we are able to investigate how many antennas we need to ensure a certain coverage requirement.
	
	\emph{Remark 5:} Note that the manipulation in Corollary \ref{coro1} is based on the proposed analytical framework in Section \ref{sec_frame}. Especially, it benefits greatly from the delicate tackling of the gamma distributed signal power, via a lower triangular Toeplitz matrix representation. If the upper bound in \cite{mine, venugopal2015interference,robertadhoc,robertcoverage} was used instead, we would not be able explicitly reveal the impact of antenna arrays, which, from another perspective, confirms the advantages of the proposed analytical framework.
	
	\section{Coverage Analysis for mm-Wave Cellular Networks}\label{sec_cellular}
	In this section, we will analyze the coverage probability for mm-wave cellular networks. While the sinc antenna pattern can still be employed to get a highly accurate approximation of the actual antenna pattern, its numerical evaluation is more complicated and the expression reveals little insight.
	In particular, as \eqref{frameexpr} showed, an additional integral is needed over the distance between the serving BS and the typical user. Furthermore, since $\kappa=r_0$ in cellular networks, the summation of infinite terms in the integrand does not converge quickly.
	Instead, we will analyze the cosine antenna pattern in this section, which will provide a more tractable expression. The impact of antenna arrays on the coverage probability will then be investigated.
	
	\subsection{Coverage Analysis}
	%As stated in Section \ref{II-A}, one major difference between ad hoc and cellular networks is the spatial distribution of the interferers, i.e., the distribution of the interfering BSs $\Phi\backslash\{x_0\}$ in cellular networks is a conditional PPP given $x_0$, within a ring with internal diameter $r_0$ and external diameter $R$. This fact will unfortunately lead to additional difficulties in numerical evaluation if we still adopt the sinc antenna pattern. This motivates us to consider another approximate antenna pattern that maintains a good balance between the accuracy and tractability in coverage analysis for mm-wave cellular networks, i.e., the cosine antenna pattern. 
	In contrast to existing works \cite{robertcoverage}, we will present an analytical result for coverage probability that fully reflects the directionality in mm-wave cellular networks. Note that although the proposed approximate antenna pattern is more complicated than the flat-top pattern, the new expression based on the analytical framework in Section \ref{sec_frame} is more compact and tractable. With the cosine antenna pattern, the coverage probability is derived in the following proposition.
	\begin{prop}\label{th3}
		The coverage probability of mm-wave cellular networks with the cosine antenna pattern is given by
		\begin{equation}
			p_\mathrm{c}^{\cos}(\tau)=\pi\lambda_\mathrm{b}\int_0^{R^2}e^{-\pi\lambda_\mathrm{b}r}\left\Vert\exp\left\{\frac{1}{N_\mathrm{t}}\mathbf{C}_M(r)\right\}\right\Vert_1\mathrm{d}r.\label{26}
		\end{equation}
		The nonzero entries in $\mathbf{C}_M$ are determined by
		\begin{equation}
			\begin{split}
				c_k(r)&=\frac{2\sqrt{\pi}\lambda_\mathrm{b}\lambda\Gamma\left(k+\frac{1}{2}\right)\Gamma(M+k)\tau^k}{d(k!)^2(\alpha k-2)\Gamma(M)}\\
				&\relphantom{=}\times\left[J_k\left(-\tau\right)r-J_k\left(-\frac{\tau}{R^\alpha}r^\frac{1}{\delta}\right)R^{2-\alpha k}r^\frac{ k}{\delta}\right]\\
				&\relphantom{=}+\mathbf{1}(k\le1)\frac{(-1)^{k+1}M\tau\sigma^2}{\beta P_\mathrm{t}},
			\end{split}
		\end{equation}
		where
		\begin{equation}
			J_k\left(x\right)={}_3F_2\left(k+\frac{1}{2},k-\delta,k+M;k+1,k+1-\delta;x\right),
		\end{equation}
		with ${}_3F_2(a_1,a_2,a_3;b_1,b_2;z)$ denoting the generalized hypergeometric function.
	\end{prop}
	\begin{IEEEproof}
		See Appendix \ref{AF}.
	\end{IEEEproof}
	
	Note that the coefficients $c_k(r)$ in Proposition \ref{th3} can be expressed based on the well-known hypergeometric function rather than the infinite summations as in Proposition \ref{th2} for ad hoc networks, which are efficiently calculated in modern numerical software. This illustrates that the cosine antenna pattern enables a more tractable analysis for cellular networks. We will use Proposition \ref{th2} as an approximation for the coverage probability with the actual antenna pattern, while the accuracy of the cosine antenna pattern will be verified in Section \ref{numer}. Furthermore, similar to Remark 3, the swap of two expectations operated in \eqref{swap} enables the derivation of Proposition \ref{th3} and turns out to be an effective and tractable approach to tackle complicated channel gain distributions in mm-wave networks.
	
	\emph{Remark 6:} With Proposition \ref{th3}, we can numerically calculate the required BS density as well as the minimum number of antennas for a desirable coverage probability. Furthermore, the optimal BS density that achieves the maximum coverage probability, as we will see in Section \ref{VI-A}, can also be numerically determined by Proposition \ref{th3}.
	
	\subsection{Impact of Directional Antenna Arrays}
	In the last subsection, we have derived an analytical result for coverage probability of mm-wave cellular networks with the cosine antenna pattern. However, it is difficult to further analyze the impact of directional antenna arrays since there is an extra integral of the induced $\ell_1$-norm of the matrix exponential, which contains the array size parameter $N_\mathrm{t}$. As an alternative, a lower bound for the coverage probability in Proposition \ref{th3} is provided next, based on which we will present the impact directional antenna arrays.
	\begin{corollary}\label{coro2}
		A lower bound of the coverage probability \eqref{26} is given by
		\begin{equation}\label{cecov}
			p_\mathrm{c}^{\cos}(\tau)\ge\left(1-e^{-\pi\lambda_\mathrm{b}R^2}\right)\left\Vert\exp\left\{\frac{1}{N_\mathrm{t}(1-e^{-\pi\lambda_\mathrm{b}R^2})}\mathbf{D}_M\right\}\right\Vert_1.
		\end{equation}
		The nonzero entries in $\mathbf{D}_M$ determined by
		\begin{equation}
		\begin{split}
			d_k&=\frac{2\lambda\Gamma\left(k+\frac{1}{2}\right)\Gamma(M+k)\tau^k}{\sqrt{\pi}d(k!)^2(\alpha k-2)\Gamma(M)}
			\bigg[y_k\left(-\tau\right)-(\pi\lambda_\mathrm{b})^2R^{2-\alpha k}\\
			&\relphantom{=}\times\int_0^{R^2}e^{-\pi\lambda_\mathrm{b}r}r^\frac{\alpha k}{2}J_k\left(-\frac{\tau}{R^{\alpha}}r^\frac{1}{\delta}\right)\mathrm{d}r\bigg]\\
			&\relphantom{=}+\mathbf{1}(k\le1)\frac{(-1)^{k+1}M\tau\sigma^2}{\beta P_\mathrm{t}\left({\pi\lambda_\mathrm{b}}\right)^{\frac{1}{\delta}}}\gamma\left(1+\frac{1}{\delta},\pi\lambda_\mathrm{b}R^2\right),
		\end{split}
		\end{equation}
		where $\gamma(s,x)$ is the lower incomplete gamma function \cite[Page 890]{zwillinger2014table} and 
		\begin{equation}
			\begin{split}
			y_k(x)&=J_k(x)\left[1-e^{-\pi\lambda_\mathrm{b}R^2}\left(1+\pi\lambda_\mathrm{b}R^2\right)\right]\\
			&\relphantom{=}+\mathbf{1}(k=0)\left(
			\pi\lambda_\mathrm{b}R^2-1+e^{-\pi\lambda_\mathrm{b}R^2}\right).
			\end{split}
		\end{equation}
	\end{corollary}
	\begin{IEEEproof}
		See Appendix \ref{AG}.
	\end{IEEEproof}
	
	With this lower bound, the integrand no longer involves the induced $\ell_1$-norm of a matrix exponential, which creates the possibility to disclose the impact of antenna arrays as stated in the following corollary.
	\begin{corollary}\label{coro3}
		The lower bound of the coverage probability \eqref{26} is a non-decreasing concave function of the array size, and it can be rewritten as
		\begin{equation}\label{b33}
			p_\mathrm{c}^{\cos}(t)\ge \left(1-e^{-\pi\lambda_\mathrm{b}R^2}\right)e^{\beta_0t}\left(1+\sum_{n=1}^{M-1}\beta_nt^n\right),
		\end{equation}
		where $t=\frac{1}{N_\mathrm{t}}$ is the inverse of the array size and
		\begin{equation}
			\beta_n=
			\begin{dcases}
				\frac{d_0}{1-e^{-\pi\lambda_\mathrm{b}R^2}}&n=0,\\
				\frac{\left\Vert \left(\mathbf{D}_M-d_0\mathbf{I}_M\right)^n\right\Vert_1}{n!\left(1-e^{-\pi\lambda_\mathrm{b}R^2}\right)}&n\ge1.
			\end{dcases}
		\end{equation}
		When $t\to0$, i.e., $N_\mathrm{t}\to\infty$, the asymptotic outage probability is given by
		\begin{equation}
			\tilde{p}_\mathrm{o}^\mathrm{cos}(t)\sim \frac{\mu}{N_\mathrm{t}}+e^{-\pi\lambda_\mathrm{b}R^2},
		\end{equation}
		where $\mu=-\sum_{n=0}^{M-1}d_n>0$. 
	\end{corollary}
	\begin{IEEEproof}
		The proof is similar to that of Corollary \ref{coro1}.
	\end{IEEEproof}
	It turns out that this lower bound of the coverage probability with the array size is quite similar to that in mm-wave ad hoc networks, yet with additional terms brought by the user association. This similarity shows that the impact of directional antenna arrays in mm-wave networks does not depend much on the user association strategy. Although this result is based on the cosine antenna pattern and a lower bound, later we will show its accuracy via simulations. Similar to Remark 5, the key tool here is the analytical framework proposed in Section \ref{sec_frame}, which enables us to investigate the impact of antenna arrays in mm-wave cellular networks.  
	
	\section{Numerical Results}\label{numer}
		\begin{figure*}
		\centering
		\subfigure[SINR, SIR, and SNR coverage probabilities in mm-wave cellular networks when $R=200$ m, $N_\mathrm{t}=64$, $\tau=10$ dB, $M=3$, and $\alpha=2.1$.]
		{
			\centering\includegraphics[height=5.6cm]{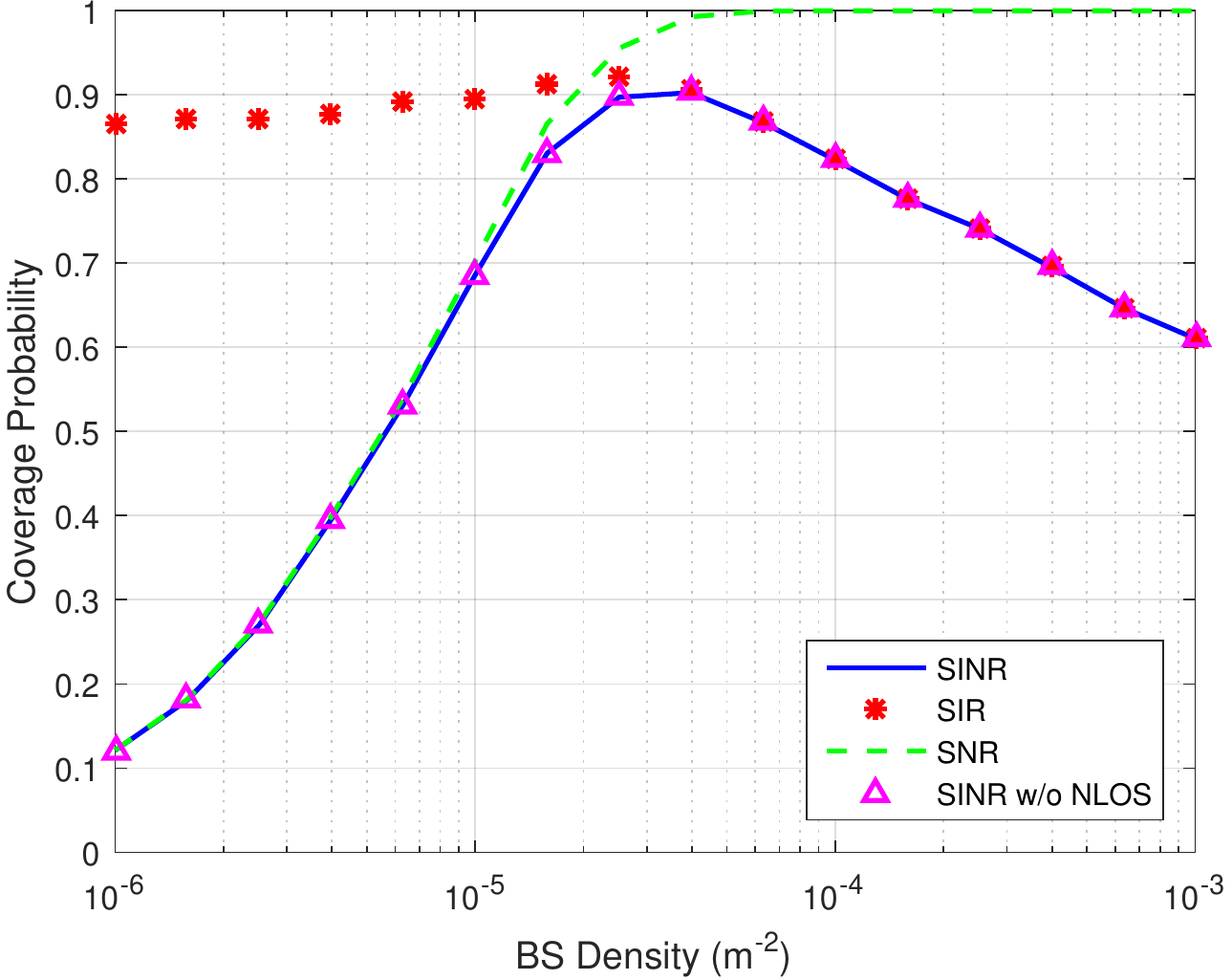}\label{f1a}
		}
		\subfigure[SINR, SIR, and SNR coverage probabilities in mm-wave ad hoc networks when $R=180$ m, $N_\mathrm{t}=64$, $\tau=5$ dB, $M=5$, $\alpha=2.2$, and $r_0=25$ m.]{
			\centering\includegraphics[height=5.6cm]{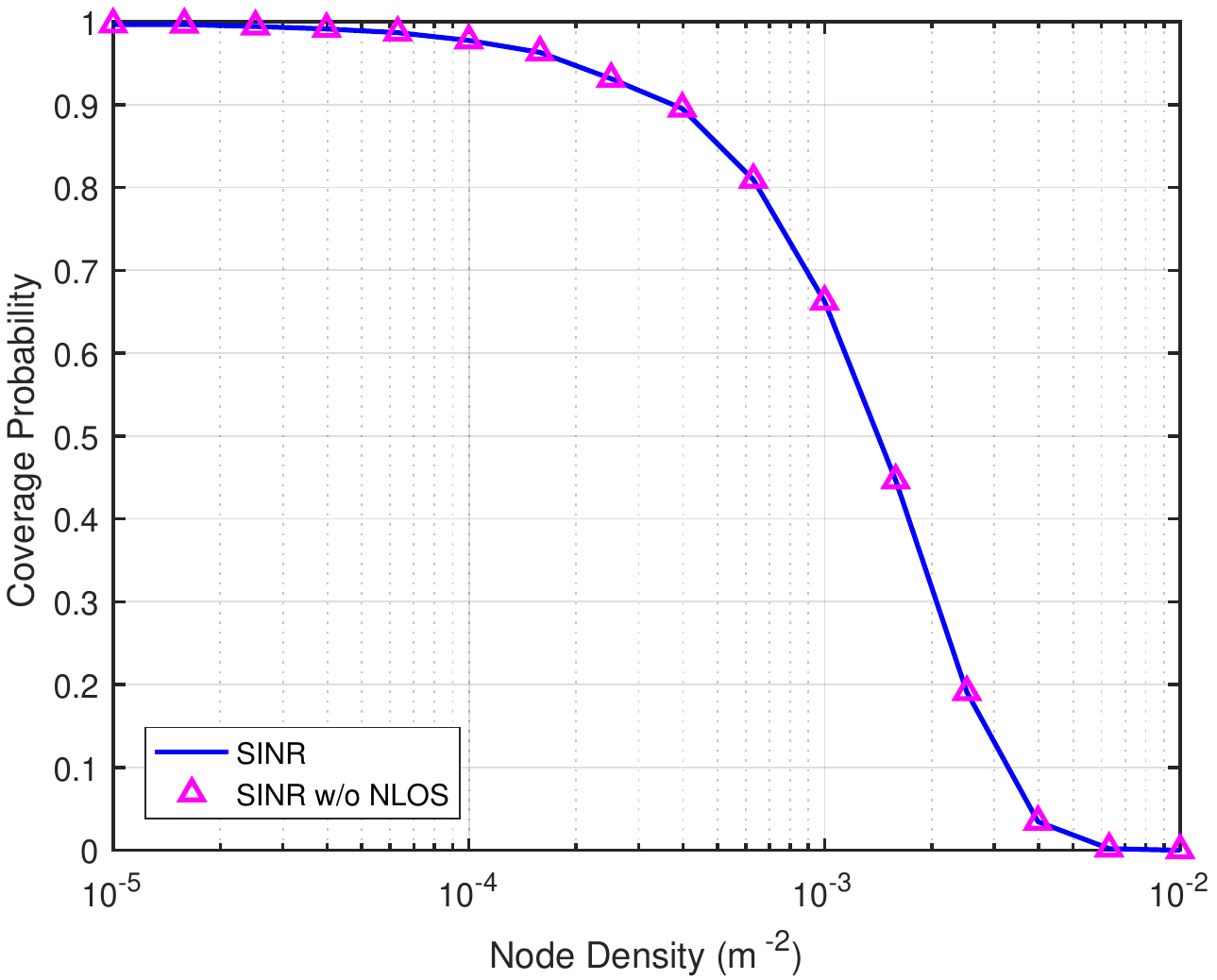}\label{f1b}
		}
		\caption{The impact of NLOS signals and interference in mm-wave (a) cellular networks and (b) ad hoc networks.}\label{f1}
	\end{figure*}

	In this section, we will present numerical results of coverage probabilities in both mm-wave ad hoc and cellular networks. We assume that the bandwidth is 1 GHz, and the transmit power of each BS is set as 1 Watt. The separation between the antenna elements is $d=\frac{\lambda}{4}$, i.e., quarter-wavelength to avoid the grating lobes. From the recent measurements of mm-wave signal propagations \cite{rappaportchannel}, the path loss exponent $\alpha$ is close to 2 and the intercept is $\beta=-61.4$ dB. All simulation results shown in this section are averaged over $5\times10^5$ realizations.
	
	\subsection{The Role of NLOS Signals and Interference}\label{VI-A}
	In Section \ref{II-A}, we stated the assumption that NLOS signals and NLOS interference are negligible in mm-wave networks, which will be justified in this subsection. To model the NLOS signals and interference, we set the propagation parameters as follows: the path loss exponent is $\alpha_\mathrm{NLOS}=4$ and the intercept is $\beta_\mathrm{NLOS}=-72$ dB \cite{rappaportchannel}. Due to the richer reflections and scattering environment of NLOS propagations, Rayleigh fading is assumed as the small-scale propagation model of the NLOS signals and interference.  According to recent measurements, a practical value of the LOS ball radius $R$ should be in the order of hundred meters \cite{7593259}. 
	
	In Fig. \ref{f1a}, we show a simulation of the SINR coverage probability without incorporating the NLOS serving BS and NLOS interferers, whose curve almost coincides with that including NLOS components. This demonstrates that the impact of NLOS signals and interference is negligible and validates the LOS assumption made in Section \ref{II-A}, i.e., we only need to focus on the analysis where the typical receiver is associated with a LOS transmitter and the interference is brought by LOS interferers. The underlying reasons are as follows for different BS densities: 1) When the BS density is low, the network is operating in the noise-limited regime, and thus only LOS signal matters; 2) At medium BS densities, there is a certain probability to have a LOS serving BS, and the interference gradually affects the SINR coverage. However, the LOS interference power is much higher compared to the NLOS ones. On the other hand, when the typical link is NLOS, it is difficult to achieve a satisfactory SINR value; 3) Very dense mm-wave networks will be LOS interference-limited, which has been investigated in \cite{mine,robertcoverage}.
	
	For mm-wave ad hoc networks, the typical dipole pair is assumed to be LOS. As explained in Section \ref{IV-A}, the coverage probability is unsatisfactory due to the huge path loss and high noise power when the signal link is NLOS. Fig. \ref{f1b} demonstrates the impact of NLOS interferers when the tagged transmitter is in the LOS condition. It manifests that, with a LOS transmitter associated with the typical receiver, the NLOS interference is also negligible for the reasons which are similar to those in cellular networks \cite{robertadhoc}.
	
	Hence, it is reasonable to neglect the NLOS components in the analysis for mm-wave networks. Although we showed that NLOS parts are minor, note that all the simulations include them to maintain the completeness and consistency. Moreover, we retain the actual antenna pattern \eqref{sinsin} in the remaining simulations.
	
	\subsection{Coverage Analysis}
	\begin{figure*}
		\centering
		\subfigure[Coverage probability in mm-wave ad hoc networks.]
		{
			\centering\includegraphics[height=5.6cm]{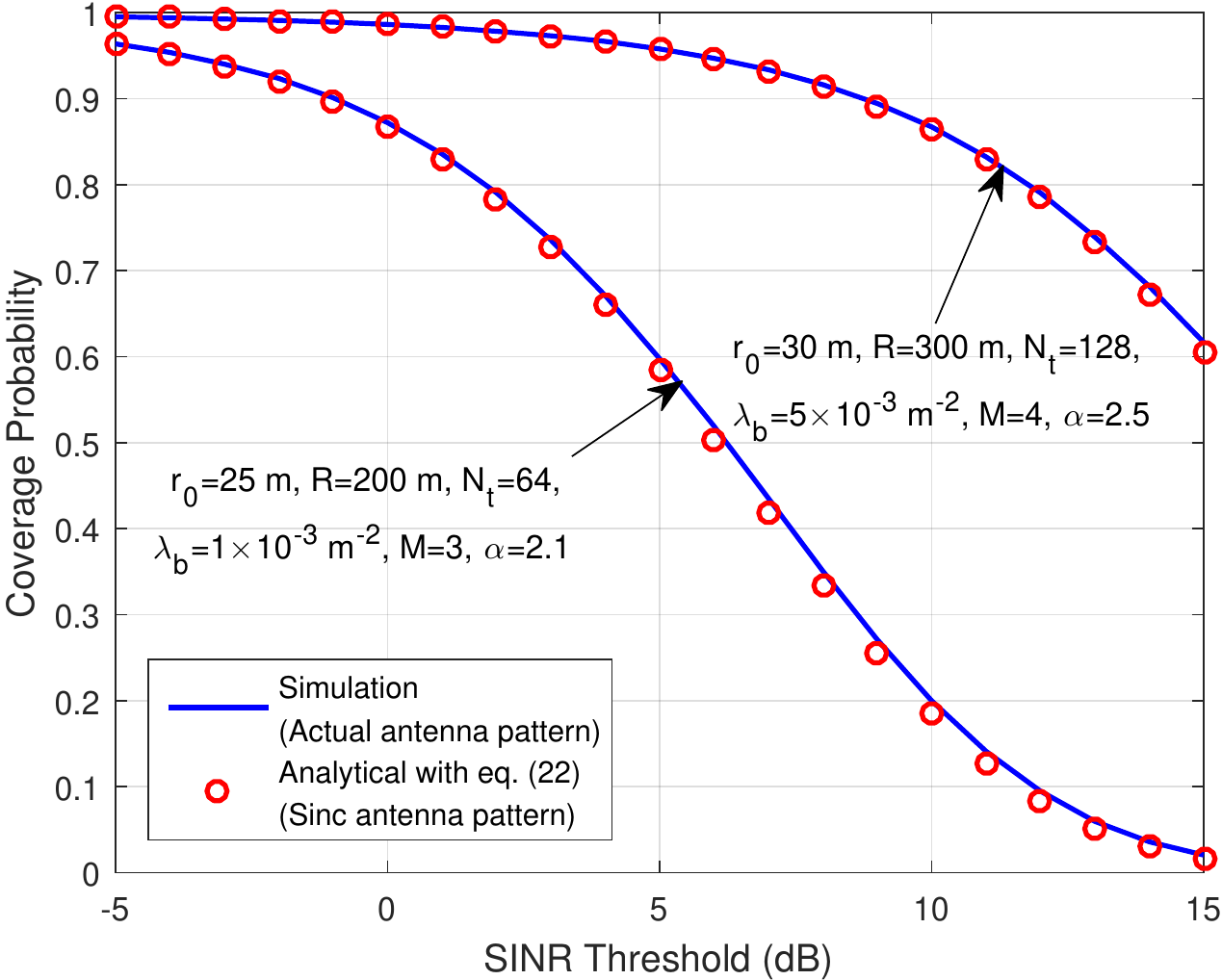}\label{f2a}
		}
		\subfigure[Coverage probability in mm-wave cellular networks when $R=200$ m, $N_\mathrm{t}=128$, $\lambda_\mathrm{b}=1\times 10^{-3}$ m$^{-2}$, $M=3$, and $\alpha=2.1$.]{
			\centering\includegraphics[height=5.6cm]{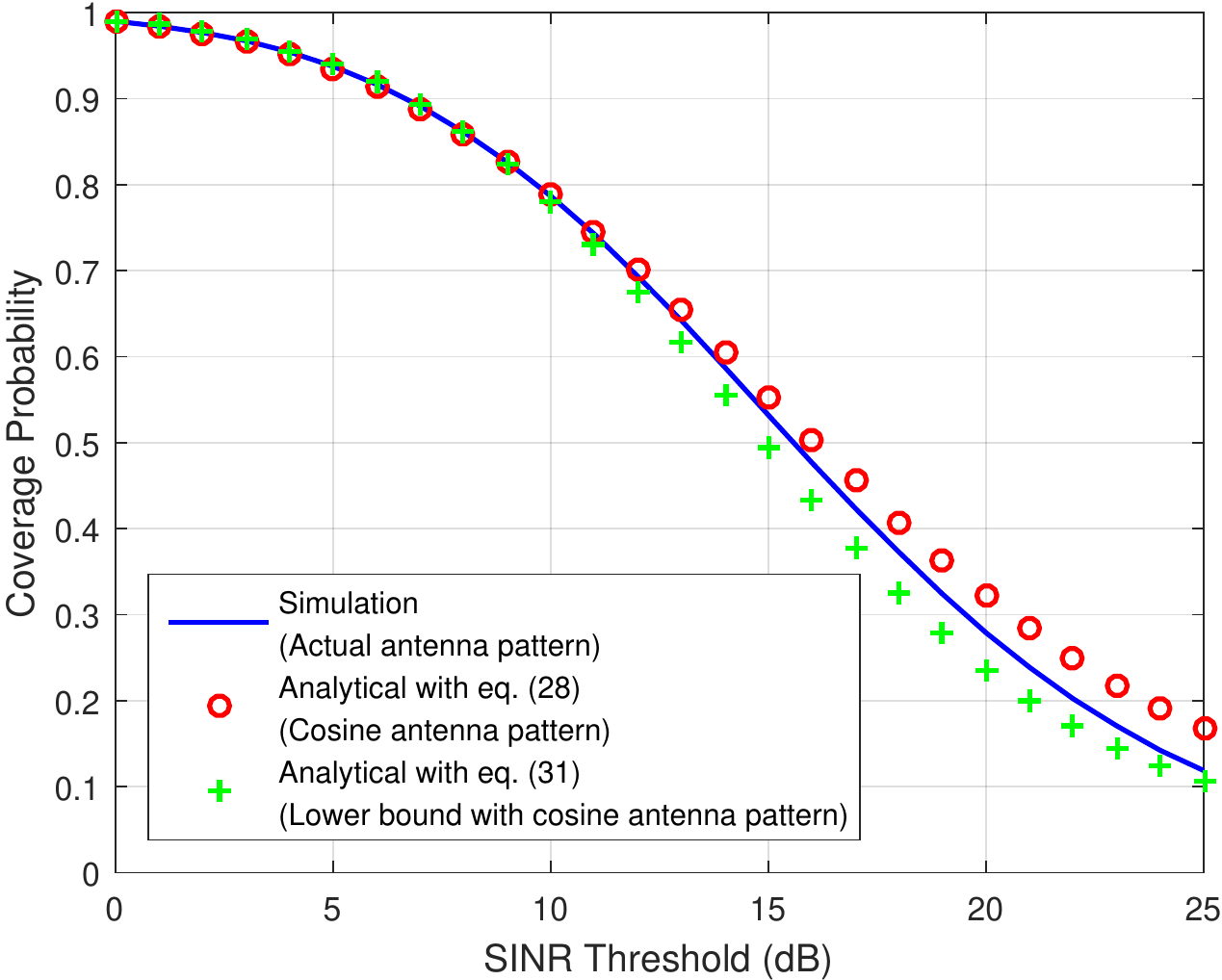}\label{f2b}
		}
		\caption{Coverage analysis using (a) Proposition \ref{th2} for mm-wave ad hoc networks, and (b) Proposition \ref{th3} and Corollary \ref{coro2} for mm-wave cellular networks.}
	\end{figure*}
	
	The effects of noise and interference in mm-wave networks are also investigated. In Fig. \ref{f1a}, we evaluate the signal-to-interference ratio (SIR) and signal-to-noise (SNR) coverage probabilities versus the BS density in mm-wave cellular networks. 
	It was found in \cite{7061455} that the SIR coverage probability will monotonically decrease with the increasing BS density in sub-6 GHz networks with the dual-slope path loss model, which, however, no longer holds in mm-wave cellular networks.
	This is because the difference in the small-scale fading for LOS and NLOS propagations in mm-wave networks, which were assumed to be the same in \cite{7061455}. 
	When the BS density gradually increases, the signal link tends to experience Nakagami fading rather than Rayleigh fading. This change in small-scale fading results in a slight increase of the SIR coverage probability, which also implicitly illustrates that Nakagami fading provides better coverage than Rayleigh fading.
	Therefore, as a lower bound of both SIR and SNR coverage probabilities, the SINR coverage probability in mm-wave cellular networks has a peak value with the increasing BS density. On the other hand, different from mm-wave cellular networks, the SINR coverage probability decreases with network densification due to the fixed dipole distance and arbitrarily close interferers, which is shown in Fig. \ref{f1b}. This evaluation indicates the importance of analyzing the SINR distribution in mm-wave cellular networks, while the SIR coverage can be used as a good metric for mm-wave ad hoc networks.

	In this subsection, we will verify our analytical results in Sections \ref{sec_ad} and \ref{sec_cellular} through simulations.
	In Fig. \ref{f2a}, the SINR coverage probabilities for mm-wave ad hoc networks are evaluated. It can be seen that the analytical results match the simulations with negligible gaps, which implies the accuracy of the bound in Proposition \ref{th2}. In Remark 1, we mentioned that using finite terms for the summations in $\{c_k\}_{k=0}^{M-1}$ is sufficient. In the numerical evaluation of Proposition \ref{th2} in Fig. \ref{f2a}, we only use 5 terms in the summations and it turns out that the higher-order terms are negligible for practical evaluation.
	
	In Fig. \ref{f2b}, the coverage probability for a mm-wave cellular network is evaluated. We see that both the analytical results in Proposition \ref{th3} and Corollary \ref{coro2} give an approximate coverage probability with minor gaps. The expression in Proposition \ref{th3} yields a very good approximation for smaller SINR thresholds and a tight bound for larger ones. This is because the major approximations made in the cosine antenna pattern \eqref{approxpattern} are on the side lobe gains that are approximated to be zeros, while the main lobe gains are approximated accurately with the cosine function. When the SINR threshold gets large, the interference power is smaller, which also means the interference is more likely to be produced by side lobe gains. Therefore, the gap will gradually increase due to the relatively crude approximation of the side lobe gains.
	
	The analytical result in Corollary \ref{coro2} provides a lower bound of the expression in Proposition \ref{th3}. Although it is not guaranteed to be a lower or upper bound of the exact SINR coverage probability, it gives a good approximation as shown in Fig. \ref{f2b}, with more analytical tractability and potential for further analysis, which will be discussed in detail in the next subsection. The results presented in Fig. \ref{f2b} show the effectiveness and rationale of the proposed cosine antenna pattern \eqref{approxpattern} in coverage analysis for mm-wave cellular networks, which is an ideal candidate for further performance analysis in mm-wave cellular networks.

	\subsection{Impact of Directional Antenna Arrays}
		\begin{figure*}
		\centering
		\subfigure[Impact of antenna arrays in mm-wave ad hoc networks when $R=200$ m, $\tau=5$ dB, $\lambda_\mathrm{b}=1\times 10^{-3}$ m$^{-2}$, $\alpha=2.1$, and $r_0=25$ m.]
		{
			\centering\includegraphics[height=5.6cm]{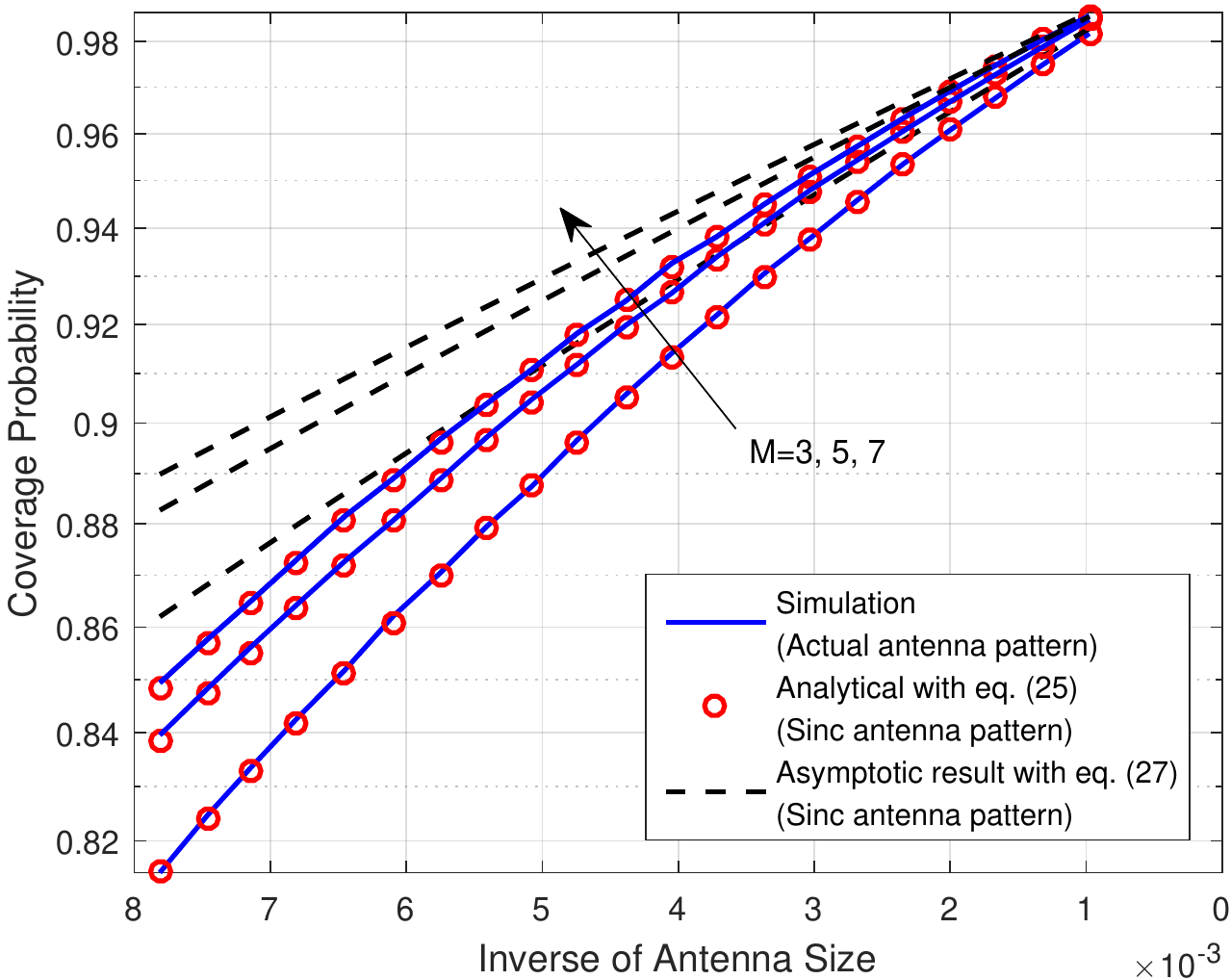}\label{f3a}
		}
		\subfigure[Impact of antenna arrays in mm-wave cellular networks when $R=200$ m, $\tau=5$ dB, $\lambda_\mathrm{b}=1\times 10^{-3}$ m$^{-2}$, and $\alpha=2.1$.]
		{
			\centering\includegraphics[height=5.6cm]{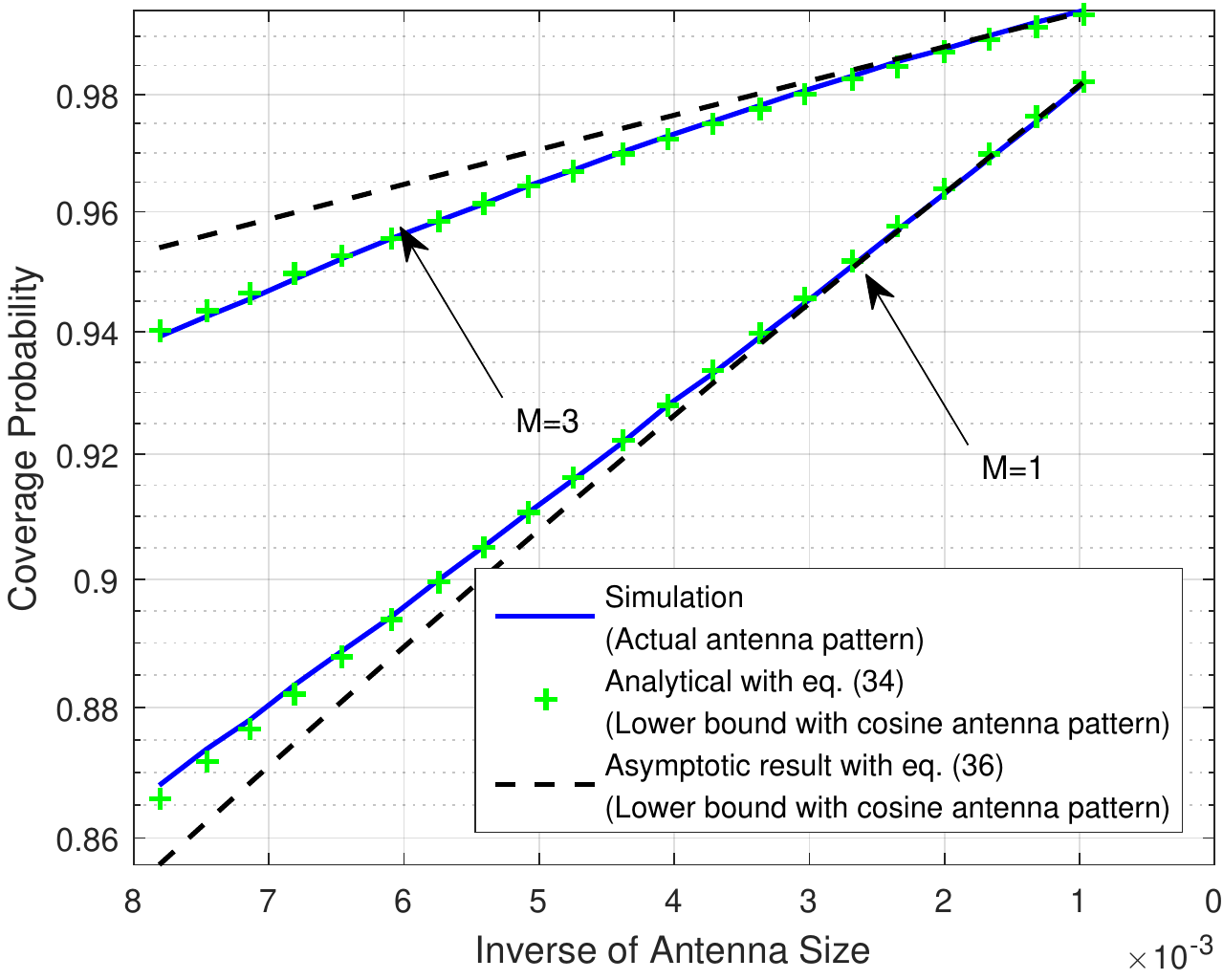}\label{f3b}
		}
		\caption{Investigation on the impact of antenna arrays using (a) Corollary \ref{coro1} for mm-wave ad hoc networks, and (b) Corollary \ref{coro3} for mm-wave cellular networks.}
	\end{figure*}
	In this subsection, we will discuss the impact of directional antenna arrays on coverage probability in mm-wave networks\footnote{In Figs. \ref{f3a} and \ref{f3b}, the x-axes are reversed, and the y-axes are in the logarithm scale.}. 
	Fig. \ref{f3a} demonstrates that the analytical result in Corollary \ref{coro1} for mm-wave ad hoc networks well matches the simulation result. We see that the increase of the array size leads to an improvement of the coverage probability, which confirms the monotonicity property in Corollary \ref{coro1}. In the following, we provide some intuitive explanations for this phenomenon. 
	The increase of the array size increases the maximum array gain for both signal and interference at the same pace, in proportion to the array size, and therefore there is almost no performance gain from increasing the maximum array gain via enlarging the array size. Nevertheless, another effect of the increasing array size for the interference is the narrowing of the beams, which reduces the probability that the interferers direct the main lobes towards the typical receiver. Moreover, note that the lower bound derived in Corollary \ref{coro1} is non-decreasing concave, which means that the benefits on the coverage from leveraging more antennas gradually diminishes with the increasing antenna size. In addition, we discover that the increase of the Nakagami parameter $M$ results in an increase of the coverage probability.
		%In addition, the impact of the small-scale fading is investigated. We discover that the increase of the Nakagami parameter $M$ results in an increase of the coverage probability. Generally, increasing the Nakagami parameter will tend to the non-fading scenario and therefore enhances both the desired signal and interference power. However, the directional antenna arrays suppress the interference power significantly, and thus the small-scale fading has a minor impact on interference in mm-wave networks. In other words, the increase of the Nakagami parameter mainly benefits the signal power and therefore improves the coverage probability.
	
	In Fig. \ref{f3b}, the impact of antenna arrays in mm-wave cellular networks is investigated. For the analytical result, we evaluate the coverage probability using the expression in Corollary \ref{coro3}, which gives a lower bound of the coverage probability adopting the cosine antenna pattern. Although Rayleigh fading, i.e., $M=1$, is only a special case for the analysis in this paper and not suitable for LOS mm-wave channels, it is valuable to examine this special case for checking the lower bound in Corollary \ref{coro3}. As stated in Section \ref{IV-B}, when $M=1$, the lower bound \eqref{b33} will reduce to an exponential one, which is linear in the logarithm scale shown in Fig. \ref{f3b}. When the Nakagami parameter $M$ increases, the polynomial term will take effect to make the lower bound to be a non-decreasing concave one. It turns out that the lower bound derived in Corollary \ref{coro3} can be regraded as an effective expression for analyzing the impact of directional antenna arrays in mm-wave cellular networks, and that the cosine antenna pattern is a satisfactory surrogate of the actual antenna pattern for tractable analysis in mm-wave networks.
	
	\section{Conclusions}\label{conclu}
	In this paper, we first proposed a general framework for coverage analysis in mm-wave networks. It was then applied to derive new tractable expressions of coverage probabilities for mm-wave ad hoc and cellular networks, where two approximate antenna patterns with good accuracy and analytical tractability were adopted. 
	We have shown that, as the network density increases, the coverage probability reaches a peak in mm-wave cellular networks, while it monotonically decreases in ad hoc networks.
	More importantly, analytical results show that the coverage probabilities of both types of networks increase as a non-decreasing concave function with the antenna array size.
	It will be interesting to extend the proposed analytical framework to more advanced precoding techniques, e.g., hybrid precoding \cite{el2014spatially,7397861}. Moreover, a coverage analysis that includes the beam misalignment caused by the imperfect channel information also is a promising future research direction.
	
	\appendices
	\section{}\label{AB}
	Defining $x_n=\frac{(-s)^n}{n!}\mathcal{L}^{(n)}(s)$, the coverage probability \eqref{nthde} can be expressed as
	\begin{equation}\label{25}
		p_\mathrm{c}(\tau)=\mathbb{E}_{r_0}\left[\sum_{n=0}^{M-1}x_n\right],
	\end{equation}
	where $x_0=\mathcal{L}(s)=\exp\{\eta(s)\}$ is given in Lemma \ref{lem1}. Next, we will express $x_n$ in a recursive form. It is obvious that $\mathcal{L}^{(1)}(s)=\eta^{(1)}(s)\mathcal{L}(s)$, and according to the formula of Leibniz for the $n$-th derivative of the product of two functions \cite{roman1980formula}, we have
	\begin{equation}
		\mathcal{L}^{(n)}(s)=\frac{\mathrm{d}^{n-1}}{\mathrm{d}s}\mathcal{L}^{(1)}(s)=\sum_{i=0}^{n-1}{{n-1}\choose i} \eta^{(n-i)}(s)\mathcal{L}^{(i)}(s),
	\end{equation}
	followed by
	\begin{equation}
		\frac{(-s)^n}{n!}\mathcal{L}^{(n)}(s)=\sum_{i=0}^{n-1}\frac{n-i}{n}\frac{(-s)^{(n-i)}}{(n-i)!}\eta^{(n-i)}(s)\frac{(-s)^i}{i!}\mathcal{L}^{(i)}(s).
	\end{equation}
	Therefore, the recursive relationship of $x_n$ is
	\begin{equation}\label{recur}
		x_n=\sum_{i=0}^{n-1}\frac{n-i}{n}c_{n-i}x_i,\quad c_k=\frac{(-s)^k}{k!}\eta^{(k)}(s).
	\end{equation}
	We define two power series as follows to solve for $x_n$,
	\begin{equation}\label{eq40}
		C(z)\triangleq\sum_{n=0}^\infty c_nz^n,\quad 
		X(z)\triangleq\sum_{n=0}^\infty x_nz^n.
	\end{equation}
	Following the method in \cite[Appendix A]{chang1}, using the properties that $C^{(1)}(z)=\sum_{n=0}^{\infty}nc_nz^{n-1}$ and $C(z)X(z)=\sum_{n=0}^\infty\sum_{i=0}^nc_{n-i}x_iz^n$, from \eqref{recur}, we obtain the differential equation
	\begin{equation}
		X^{(1)}(z)=C^{(1)}(z)X(z),
	\end{equation}
	whose solution is
	\begin{equation}
		X(z)=\exp\left\{C(z)\right\}.\label{40}
	\end{equation}
	Therefore, according to \eqref{25}, \eqref{eq40} and \eqref{40}, the coverage probability is given by
	\begin{equation}
	\begin{split}
		p_\mathrm{c}(\tau)&=\mathbb{E}_{r_0}\left[\sum_{n=0}^{M-1}x_n\right]=\mathbb{E}_{r_0}\left[\sum_{n=0}^{M-1}\frac{1}{n!}\left.{X^{(n)}(z)}\right|_{z=0}\right]\\
		&=\mathbb{E}_{r_0}\left[\sum_{n=0}^{M-1}\frac{1}{n!}\frac{\mathrm{d}^n}{\mathrm{d}z^n}\left.{e^{C(z)}}\right|_{z=0}\right].\label{chang}
	\end{split}
	\end{equation}
	From \cite[Page 14]{henrici1974applied}, the first $M$ coefficients of the power series $e^{C(z)}$ form the first column of the matrix exponential $\exp\{\mathbf{C}_M\}$, whose exponent is given in \eqref{topmatrix}. 
	\begin{figure*}
		\begin{equation}\label{eq45}
		c_k=\begin{dcases}
		s\sigma_\mathrm{n}^2+\delta\pi\lambda_\mathrm{b}s\left\{R^{2-\alpha}\mathbb{E}_{g}[g\mathrm{E}_{\delta}(sR^{-\alpha}g)]-\kappa^{2-\alpha}\mathbb{E}_{g}[g\mathrm{E}_{\delta}(s\kappa^{-\alpha}g)]\right\}&k=1,\\
		\frac{\pi\delta\lambda_\mathrm{b}s^k}{ k!}\left\{R^{2-\alpha k} \mathbb{E}_{g}\left[g^k\mathrm{E}_{1+\delta-k}(sR^{-\alpha}g)\right]-\kappa^{2-\alpha k}\mathbb{E}_{g}\left[g^k\mathrm{E}_{1+\delta-k}(s\kappa^{-\alpha} g)\right]\right\}&k\ge2.\\
		\end{dcases}
		\end{equation}
		\hrule
	\end{figure*}
	
	Equation \eqref{chang} can be further expressed as \eqref{frameexpr}. Furthermore, due to the fact $\frac{\mathrm{d}}{\mathrm{d}z}\mathrm{E}_p(z)=-\mathrm{E}_{p-1}(z)$, the coefficients can be recast as \eqref{eq45}.
	It can be proved that $z^{2-\alpha k}\mathrm{E}_{1+\delta-k}(z)$ is a monotone decreasing function with respect to $z$, and therefore the coefficients $c_k>0$ for $k\ge1$. 
	Summing up what has been mentioned above completes the proof of Theorem \ref{th1}.
	
	\section{
		%	Proof of Proposition \ref{th2}
	}\label{AD}
	Since $\kappa=0$ in mm-wave ad hoc networks, \eqref{intlap} can be simplified as
	\begin{equation}
		\begin{split}
			&\relphantom{=}R^2-\delta R^2\mathbb{E}_{g}\left[\mathrm{E}_{1+\delta}(sR^{-\alpha}g)\right]\\
			&\overset{(b)}=R^2-\delta R^2\Bigg\{\frac{s^\delta}{R^2}\Gamma\left(-\delta\right)\mathbb{E}_{g}\left[g^{\delta}\right]+\frac{\alpha}{2}\\
			&\relphantom{=}-\sum_{p=1}^\infty\frac{(-s)^p}{R^{\alpha p}p!\left(p-\delta\right)}\mathbb{E}_{g}\left[g^p\right]\Bigg\}\\
			&=\delta R^2\sum_{p=1}^\infty\frac{(-s)^p}{R^{\alpha p}p!\left(p-\delta\right)}\mathbb{E}_{g}\left[g^p\right]
			-\delta s^\delta\Gamma\left(-\delta\right)\mathbb{E}_{g}\left[g^{\delta}\right]\\
			&=\delta R^2\sum_{p=1}^\infty\frac{(-s)^p\Gamma(M+p)}{R^{\alpha p}p!\left(p-\delta\right)\Gamma(M)M^p}\int_0^1\frac{\sin^{2k}\left(\frac{\pi d }{\lambda}N_\mathrm{t}\theta\right)}{\left(\frac{\pi d}{\lambda}N_\mathrm{t}\theta\right)^{2k}}\mathrm{d}\theta\\
			&\relphantom{=}-\delta s^\delta\Gamma\left(-\delta\right)\frac{\Gamma\left(M+\delta\right)}{\Gamma(M)M^\delta}\int_0^1\left|\frac{\sin\left(\frac{\pi d }{\lambda}N_\mathrm{t}\theta\right)}{\frac{\pi d}{\lambda}N_\mathrm{t}\theta}\right|^{2\delta}\mathrm{d}\theta\\
			&\overset{(c)}\le \frac{R^2\lambda}{\alpha dN_\mathrm{t}}\sum_{p=1}^\infty\frac{(-s)^p{{2p-1}\bangle{p-1}} \Gamma(M+p)}{R^{\alpha p}(2p-1)!p!\left(p-\delta\right)\Gamma(M)M^p}\\
			&\relphantom{=}-\frac{\delta s^{\delta}\lambda}{\pi  dN_\mathrm{t}}\Gamma\left(-\delta\right)\frac{\Gamma\left(M+\delta\right)}{\Gamma(M)M^\delta}\xi,
		\end{split}
	\end{equation}
	where $(b)$ is from the series expansion of the generalized exponential integral. Step $(c)$ follows Lemma
	\ref{lem3}, and the upper bound is derived by extending the integral upper limit to infinity given that, for the tiny ripple tails of the $2k$-th power of the sinc, the additional integration values are extremely small and thus the upper bound in $(c)$ is tight. Therefore, the exponent of the Laplace transform is given by
	\begin{equation}
	\begin{split}
		\eta(s)&=-\frac{\pi R^2\lambda_\mathrm{b}\lambda}{\alpha dN_\mathrm{t}}\sum_{p=1}^\infty\frac{(-s)^p{{2p-1}\bangle{p-1}} \Gamma(M+p)}{R^{\alpha p}(2p-1)!p!\left(p-\delta\right)\Gamma(M)M^p}\\
		&+\frac{\delta s^{\delta}\lambda_\mathrm{b}\lambda}{  dN_\mathrm{t}}\Gamma\left(-\delta\right)\frac{\Gamma\left(M+\delta\right)}{\Gamma(M)M^\delta}\xi-\frac{s\sigma^2}{\beta P_\mathrm{t}N_\mathrm{t}}.\label{lsad}
		\end{split}
	\end{equation}
	The coefficients in Proposition \ref{th2} can be easily obtained via taking the $k$-th derivative of \eqref{lsad}.
	
	\section{
		%	Proof of Corollary \ref{coro1}
	}\label{AE}
	According to \eqref{Ls}, the Laplace transform of noise and interference is
	\begin{equation}
	\begin{split}
		\mathcal{L}(s)&=x_0=\exp\{\eta(s)\}\\
		&=\exp\left(-s\sigma_\mathrm{n}^2-\pi\lambda_\mathrm{b}R^2\left\{1-\delta \mathbb{E}_{g}\left[\mathrm{E}_{1+\delta}(sR^{-\alpha}g)\right]\right\}\right).
	\end{split}
	\end{equation}
	Note that $1-\delta \mathbb{E}_{g}\left[\mathrm{E}_{1+\delta}(sR^{-\alpha}g)\right]$ is a positive term due to the facts that $\mathrm{E}_{1+\delta}(z)$ is a monotone decreasing function of $z$ and $\mathrm{E}_{1+\delta}(0)=\frac{1}{\delta}$. Hence, the Laplace transform $x_0$ is non-decreasing with the antenna array size $N_\mathrm{t}$, where $\eta(s)$ is given in \eqref{lsad}. According to the recursive relationship \eqref{recur} between $x_n$, it turns out that every $x_n$ is a non-decreasing function of $N_\mathrm{t}$. Recalling that $p_\mathrm{c}(\tau)=\mathbb{E}_{r_0}\left[\sum_{n=0}^{M-1}x_n\right]$, the monotonicity in Corollary \ref{coro1} has been proved, and the concavity of the lower bound can be proved via similar steps.
	
	We first write $\mathbf{C}_M$ in the form
	\begin{equation}
		\mathbf{C}_M=c_0\mathbf{I}_M+(\mathbf{C}_M-c_0\mathbf{I}_M),
	\end{equation}
	where the first term is a scalar matrix. Since $\mathbf{C}_M$ is a lower triangular Toeplitz matrix, the second part is a nilpotent matrix, i.e., $(\mathbf{C}_M-c_0\mathbf{I}_M)^n=\mathbf{0}$ for $n\ge M$. Hence, according to the properties of matrix exponential, we have
	\begin{equation}
		\exp\left\{\frac{1}{N_\mathrm{t}}\mathbf{C}_M\right\}=e^{c_0\frac{1}{N_\mathrm{t}}}
		\sum_{n=0}^{M-1}\frac{1}{n!}\left[\frac{1}{N_\mathrm{t}}\left(\mathbf{C}_M-c_0\mathbf{I}_M\right)\right]^n.
	\end{equation}
	Since Theorem \ref{th1} has shown that $c_k>0$ for $k\ge1$, $\mathbf{C}_M-c_0\mathbf{I}_M$ is a strictly lower triangular Toeplitz matrix with all positive entries, and so are the matrices $(\mathbf{C}_M-c_0\mathbf{I}_M)^n$. Therefore,
	\begin{equation}
		\left\Vert\exp\left\{\frac{1}{N_\mathrm{t}}\mathbf{C}_M\right\}\right\Vert_1=
		e^{c_0\frac{1}{N_\mathrm{t}}}
		\sum_{n=0}^{M-1}\frac{1}{n!}\left[\frac{1}{N_\mathrm{t}^n}\left\Vert\left(\mathbf{C}_M-c_0\mathbf{I}_M\right)^n\right\Vert_1\right],
	\end{equation}
	which completes the proof of Corollary \eqref{eqcoro1}. 
	When $t\to0$, by omitting the higher order terms, the linear Taylor expansion of the coverage is 
	\begin{equation}
		\left\Vert\exp\left\{\frac{1}{N_\mathrm{t}}\mathbf{C}_M\right\}\right\Vert_1\sim1+\frac{c_0+\left\Vert\mathbf{C}_M-c_0\mathbf{I}_M\right\Vert_1}{N_\mathrm{t}}=1+\frac{\sum_{n=0}^{M-1}c_n}{N_\mathrm{t}},
	\end{equation}
	where the slope
	\begin{equation}
		\sum_{n=0}^{M-1}c_n\overset{(d)}{<}\sum_{n=0}^\infty c_n=\sum_{n=0}^\infty \frac{(-s)^n}{n!}\eta^{(n)}(s)\overset{(e)}{=}\eta(0)=0.
	\end{equation}
	Step $(d)$ follows the fact that $c_k>0$ for $k\ge1$ as proved in Appendix \ref{AB}, and $(e)$ follows from the Taylor expansion of $\eta(0)$ at point $s$.
	
	\section{
		%	Proof of Proposition \ref{th3}
	}\label{AF}
	Following similar steps as in Appendix \ref{AD}, \eqref{intlap} can be derived as
	\begin{equation}\label{44}
	\begin{split}
		&\relphantom{=}2\int_{r_0}^R{\left(1-\mathbb{E}_g[\exp(-sgx^{-\alpha})]\right)}x\mathrm{d}x\\
		&=\delta R^2\sum_{k=1}^\infty\frac{(-sR^{-\alpha})^k}{k!(k-\delta)}\mathbb{E}_g[g^k]-\delta r_0^2\sum_{k=1}^\infty\frac{(-sr_0^{-\alpha})^k}{k!(k-\delta)}\mathbb{E}_g[g^k].
		\end{split}
	\end{equation}
	Based on the cosine antenna pattern \eqref{approxpattern}, we have
	\begin{equation}\label{45}
		\begin{split}
			&\relphantom{=}\sum_{k=1}^\infty\frac{(-z)^k}{k!(k-\delta)}\mathbb{E}_g[g^k]\\
			&=\frac{\lambda}{\pi dN_\mathrm{t}}\sum_{k=0}^\infty\frac{(-z)^k}{k!(k-\delta)}\int_0^\pi\cos^{2k}\frac{x}{2}\mathrm{d}x+\frac{\lambda}{\delta dN_\mathrm{t}}\\
			&=\frac{\lambda}{\sqrt{\pi} dN_\mathrm{t}}\sum_{k=0}^\infty\frac{(-z)^k\Gamma\left(\frac{1}{2}+k\right)}{(k!)^2(k-\delta)}+\frac{\lambda}{\delta dN_\mathrm{t}}\\
			&\overset{(f)}{=}\frac{\lambda}{\delta dN_\mathrm{t}}\left[1-{}_3F_2\left(\frac{1}{2},-\delta,M;1,1-\delta;-\frac{z}{M}\right)\right],\\
		\end{split}
	\end{equation}
	where $(f)$ inversely applies the definition (series expansion) of the generalized hypergeometric function \cite[Page 1000]{zwillinger2014table}.
	
	Substituting \eqref{45} into \eqref{44}, the exponent of the Laplace transform is given by
	\begin{equation}
	\begin{split}
		\eta(s)&=-\frac{s\sigma^2}{\beta P_\mathrm{t}N_\mathrm{t}}-\frac{\pi\lambda_\mathrm{b}\lambda}{dN_\mathrm{t}}\bigg\{\left[J_0\left(-\frac{sr_0^{-\alpha}}{M}\right)-1\right]r_0^2\\
		&\relphantom{=}-\left[J_0\left(-\frac{s R^{-\alpha}}{M}\right)-1\right]R^2\bigg\}.
		\end{split}
	\end{equation}
	Note that the derivative for the generalized hypergeometric function is
	\begin{equation}
	\begin{split}
		&\relphantom{=}\frac{\mathrm{d}}{\mathrm{d}z}{}_3F_2(a_1,a_2,a_3;b_1,b_2;z)\\
		&=\frac{\prod_{i=1}^3a_i}{\prod_{j=1}^2b_j}{}_3F_2(a_1+1,a_2+1,a_3+1;b_1+1,b_2+1;z).
		\end{split}
	\end{equation}
	Based on this expression and Theorem \ref{th1}, the entries in $\mathbf{C}_M$ in Proposition \ref{th3} are obtained.
	
	\section{
		%	Proof of Corollary \ref{coro2}
	}\label{AG}
	Similar to Appendix \ref{AB}, we define a power series
	\begin{equation}
		Q(z)\triangleq\mathbb{E}_{r_0}\left[X(z)\right]=\sum_{n=0}^\infty q_nz^n.
	\end{equation}
	Recall that $X(z)=\exp\{C(z)\}$ in \eqref{40}, and we obtain the following lower bound with a slight abuse of notation due to the fact that $C(z)$ is a function of $r_0$ in cellular networks,
	\begin{equation}
		\begin{split}
			Q(z)&=\pi\lambda_\mathrm{b}\int_0^{R^2}\exp\left\{-\pi\lambda_\mathrm{b}r+\frac{1}{N_\mathrm{t}}C(z;r)\right\}\mathrm{d}r\\
			&=\pi\lambda_\mathrm{b}\int_0^{R^2}e^{-\pi\lambda_\mathrm{b}r}\exp\left\{\frac{1}{N_\mathrm{t}}\sum_{k=0}^\infty c_k(r)z^k\right\}\mathrm{d}r\\
			&\overset{(g)}{\ge}\left(1-e^{-\pi\lambda_\mathrm{b}R^2}\right)\exp\Bigg\{\frac{\pi\lambda_\mathrm{b}}{N_\mathrm{t}(1-e^{-\pi\lambda_\mathrm{b}R^2})}\\
			&\relphantom{=}\times\sum_{k=0}^\infty\left(\int_0^{R^2}e^{-\pi\lambda_\mathrm{b}r}c_k(r)\mathrm{d}r\right)z^k\Bigg\}\\
			&\triangleq\left(1-e^{-\pi\lambda_\mathrm{b}R^2}\right)\exp\left\{\frac{1}{N_\mathrm{t}(1-e^{-\pi\lambda_\mathrm{b}R^2})}D(z;r)\right\}.
		\end{split}
	\end{equation}
	In fact, $Q(z)$ can be viewed as $\left(1-e^{-\pi\lambda_\mathrm{b}R^2}\right)\mathbb{E}_{r_0^\prime}\left[\exp\left\{ \frac{1}{N_\mathrm{t}}C(z;r_0^\prime)\right\}\right]$ for the random variable with pdf $f_{r_0^\prime}(r)=\frac{\pi\lambda_\mathrm{b}}{1-e^{-\pi\lambda_\mathrm{b}R^2}}e^{-\pi\lambda_\mathrm{b}r}$. Due to the convexity of the exponential function, we apply Jensen's inequality in $(g)$ and obtain the lower bound.
	Therefore, the coverage probability is given by
	\begin{equation}
		p_\mathrm{c}^{\cos}(\tau)=\sum_{n=0}^{M-1}\frac{1}{n!}\frac{\mathrm{d}^n}{\mathrm{d}z^n}\left.{Q(z)}\right|_{z=0},
	\end{equation}
	which can be further expressed as in Corollary \ref{coro3}.
	%\section*{Acknowledgment}
	
	\ifCLASSOPTIONcaptionsoff
	\newpage
	\fi
	
	\bibliographystyle{IEEEtran}
	\bibliography{bare_jrnl}
	
	\begin{IEEEbiography}
		[{\includegraphics[width=1in,height=1.25in,clip,keepaspectratio]{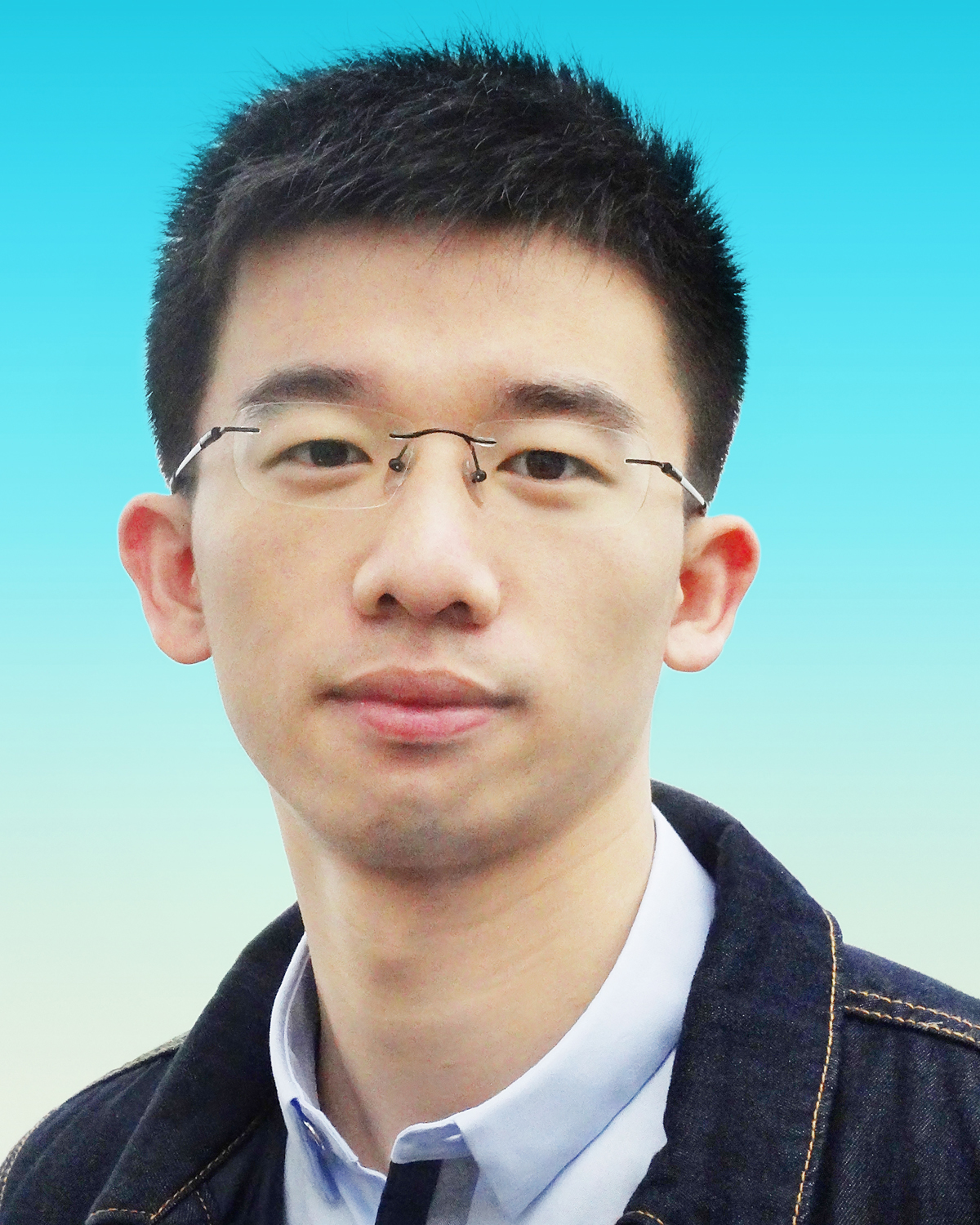}}]{Xianghao Yu}
		(S'15) received the B.Eng. degree in Information Engineering from Southeast University (SEU), Nanjing, China, in 2014. He is currently working towards the Ph.D. degree in Electronic and Computer Engineering at the Hong Kong University of Science and Technology (HKUST), under the supervision of Prof. Khaled B. Letaief. His research interests include millimeter wave communications, MIMO systems, mathematical optimization, and stochastic geometry.
	\end{IEEEbiography}

	\begin{IEEEbiography}
	[{\includegraphics[width=1in,height=1.25in,clip,keepaspectratio]{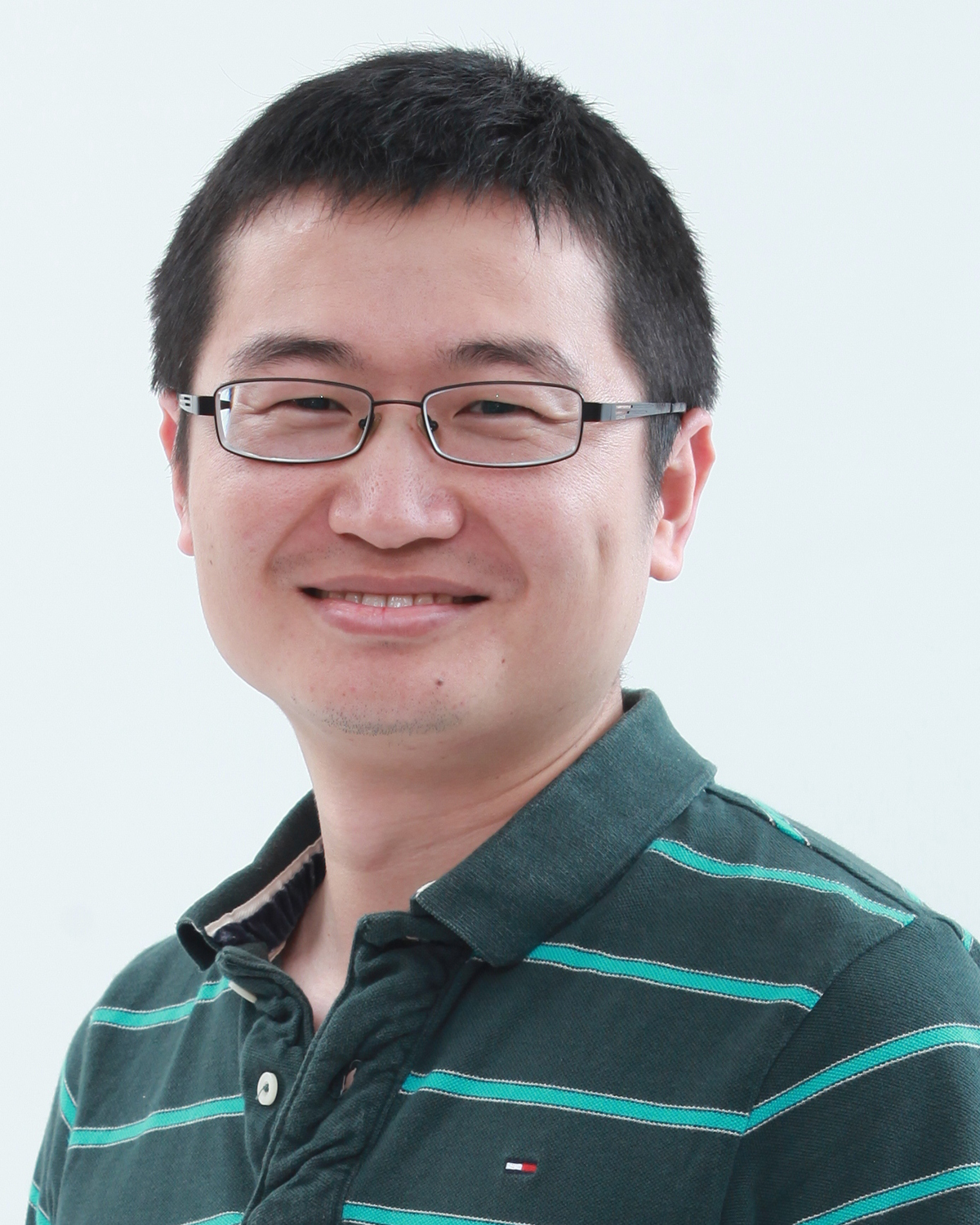}}]{Jun Zhang}
	(S'06-M'10-SM'15) received the B.Eng. degree in Electronic Engineering from the University of Science and Technology of China in 2004, the M.Phil. degree in Information Engineering from the Chinese University of Hong Kong in 2006, and the Ph.D. degree in Electrical and Computer Engineering from the University of Texas at Austin in 2009. He is currently a Research Assistant Professor in the Department of Electronic and Computer Engineering at the Hong Kong University of Science and Technology (HKUST). His current research interests include dense wireless cooperative networks, mobile edge caching and computing, cloud computing, and data analytics systems.
	
	Dr. Zhang co-authored the book \emph{Fundamentals of LTE} (Prentice-Hall, 2010). He is a recipient of the 2016 Marconi Prize Paper Award in Wireless Communications, the 2016 Young Author Best Paper Awards by the IEEE Signal Processing Society (co-author), the 2014 Best Paper Award for the \textsc{EURASIP Journal on Advances in Signal Processing}, an IEEE ICC Best Paper Award in 2016, and an IEEE PIMRC Best Paper Award in 2014. He also received the 2016 IEEE ComSoc Asia-Pacific Best Young Researcher Award. He is an Editor of \textsc{IEEE Transactions on Wireless Communications}, and is a guest editor of the special section on ``Mobile Edge Computing for Wireless Networks" in IEEE Access. He frequently serves on the technical program committees of major IEEE conferences in wireless communications, such as ICC, Globecom, WCNC, VTC, etc., and served as a MAC track co-chair for IEEE WCNC 2011.
\end{IEEEbiography}

	\begin{IEEEbiography}
	[{\includegraphics[width=1in,height=1.25in,clip,keepaspectratio]{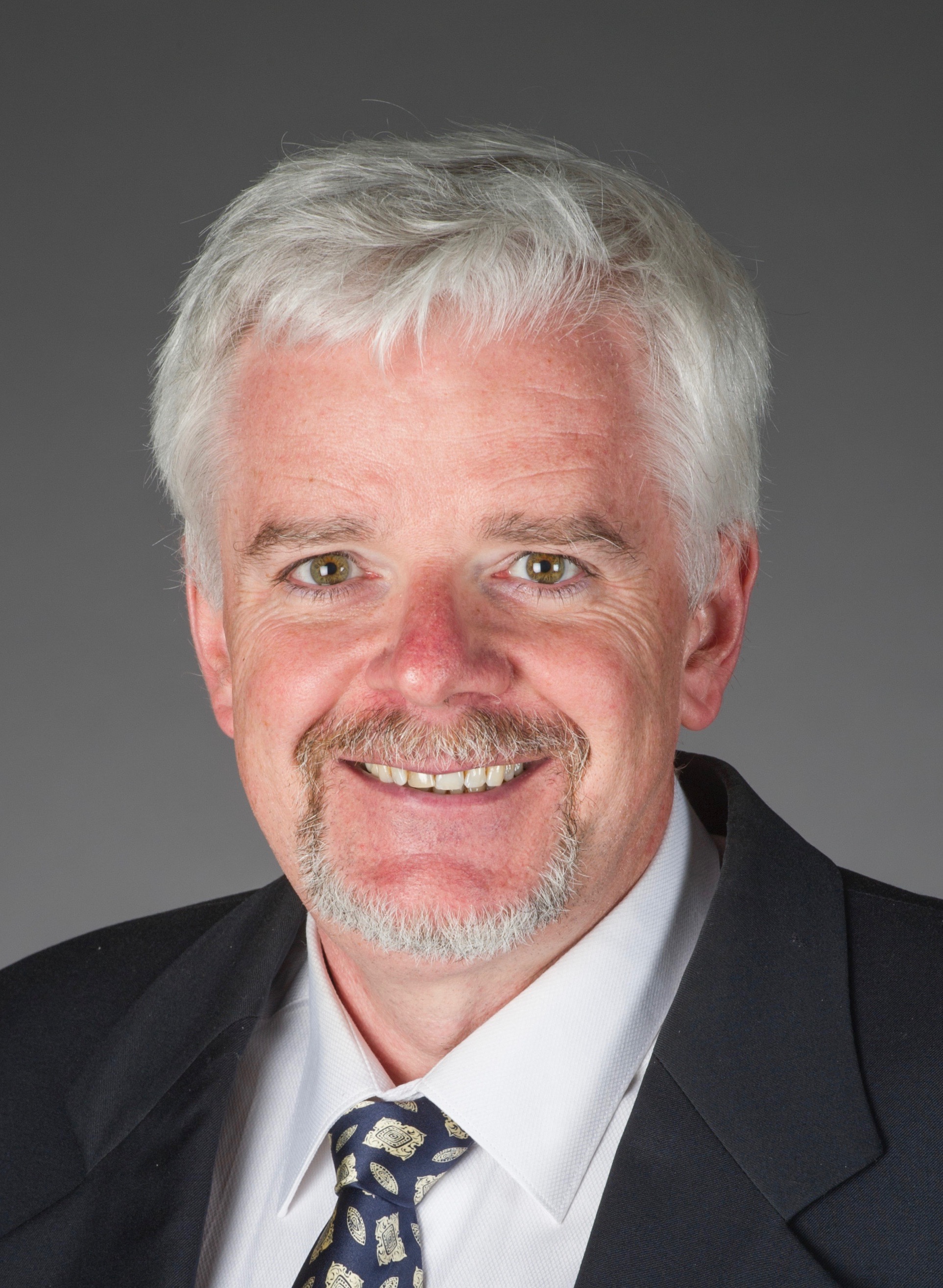}}]{Martin Haenggi}
(S'95-M'99-SM'04-F'14) received the Dipl.-Ing. (M.Sc.) and Dr.sc.techn. (Ph.D.) degrees in electrical engineering from the Swiss Federal Institute of Technology in Zurich (ETH) in 1995 and 1999, respectively. Currently he is the Freimann Professor of electrical engineering and a Concurrent Professor of applied and computational mathematics and statistics at the University of Notre Dame, Indiana, USA. In 2007-2008, he was a visiting professor at the University of California at San Diego, and in 2014-2015 he was an Invited Professor at EPFL, Switzerland.

He is a co-author of the monograph ``Interference in Large Wireless Networks" (NOW Publishers, 2009) and the author of the textbook ``Stochastic Geometry for Wireless Networks" (Cambridge University Press, 2012), and he published 14 single-author journal articles. His scientific interests include networking and wireless communications, with an emphasis on cellular, amorphous, ad hoc (including D2D and M2M), cognitive, and vehicular networks.

He served an Associate Editor of the \textsc{Elsevier Journal of Ad Hoc Networks}, the \textsc{IEEE Transactions on Mobile Computing} (TMC), the \textsc{ACM Transactions on Sensor Networks}, as a Guest Editor for the \textsc{IEEE Journal on Selected Areas in Communications}, the \textsc{IEEE Transactions on Vehicular Technology}, and the \textsc{EURASIP Journal on Wireless Communications and Networking}, as a Steering Committee member of the TMC, and as the Chair of the Executive Editorial Committee of the \textsc{IEEE Transactions on Wireless Communications} (TWC). Currently he is the Editor-in-Chief of the TWC. He also served as a Distinguished Lecturer for the IEEE Circuits and Systems Society, as a TPC Co-chair of the Communication Theory Symposium of the 2012 IEEE International Conference on Communications (ICC'12), of the 2014 International Conference on Wireless Communications and Signal Processing (WCSP'14), and the 2016 International Symposium on Wireless Personal Multimedia Communications (WPMC'16), as a General Co-chair of the 2009 International Workshop on Spatial Stochastic Models for Wireless Networks (SpaSWiN'09), and the 2012 DIMACS Workshop on Connectivity and Resilience of Large-Scale Networks, as well as a Keynote Speaker at 10 international conferences and workshops.

For both his M.Sc. and Ph.D. theses, he was awarded the ETH medal, and he received a CAREER award from the U.S. National Science Foundation in 2005 and the 2010 IEEE Communications Society Best Tutorial Paper award.
\end{IEEEbiography}

\begin{IEEEbiography}
	[{\includegraphics[width=1in,height=1.25in,clip,keepaspectratio]{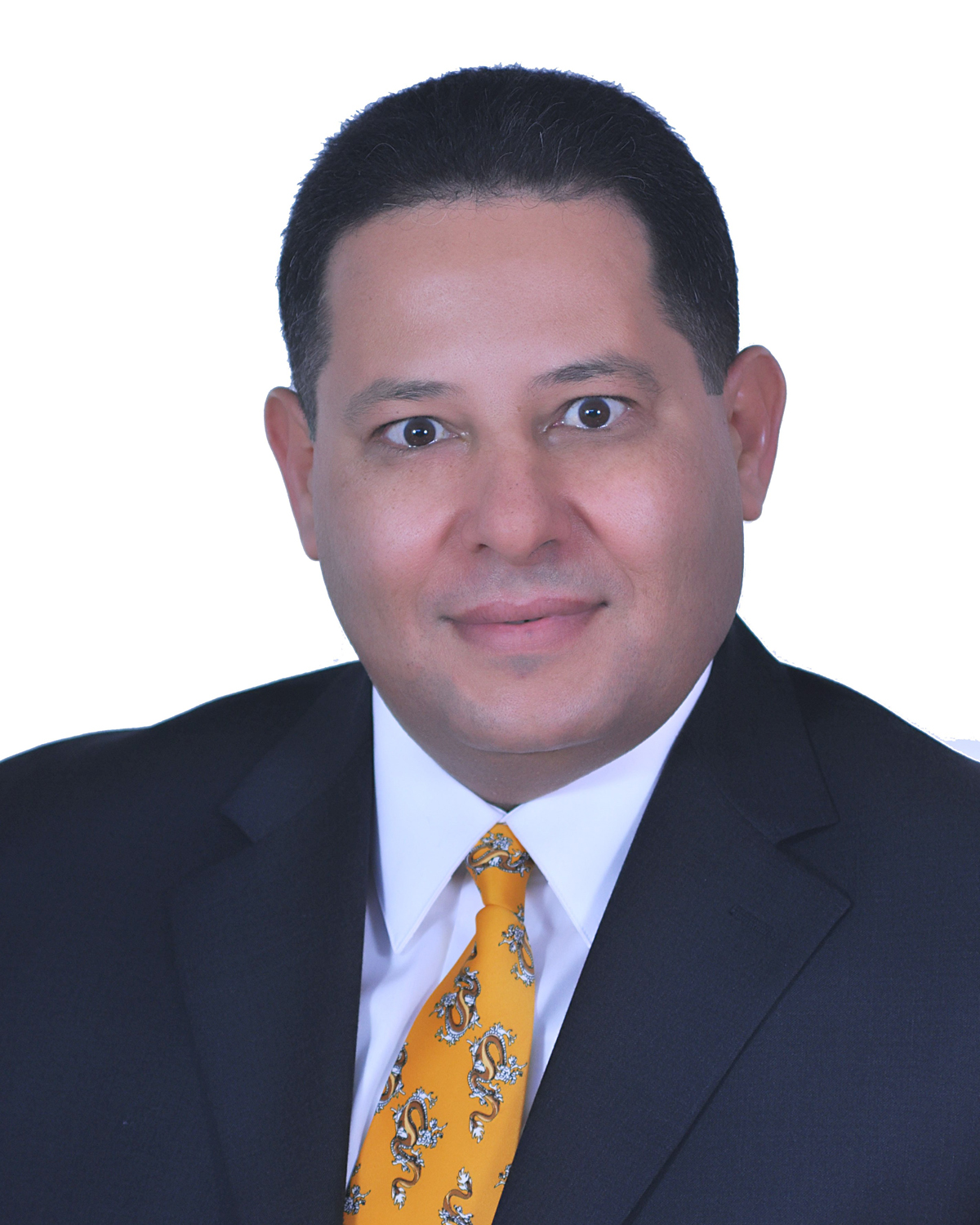}}]{Khaled B. Letaief}
	(S'85--M'86--SM'97--F'03) received the BS degree \emph{with distinction} in Electrical Engineering from Purdue University at West Lafayette, Indiana, USA, in December 1984. He received the MS and Ph.D. Degrees in Electrical Engineering from Purdue University, in August 1986, and May 1990, respectively.  
	
	From 1990 to 1993, he was a faculty member at the University of Melbourne, Australia.  He has been with the Hong Kong University of Science \& Technology since 1993 where he is known as one of HKUST’s most distinguished professors.  While at HKUST, he has held numerous administrative positions, including Head of the Electronic and Computer Engineering department, Director of Huawei Innovation Laboratory, and Director of the Hong Kong Telecom Institute of Information Technology.  While at HKUST he has also served as Chair Professor and Dean of Engineering.  Under his leadership, the School has also dazzled in international rankings (rising from \#26 in 2009 to \#14 in the world in 2015 according to QS World University Rankings.)  From September 2015, he joined HBKU as Provost to help establish a research-intensive university in Qatar in partnership with esteemed universities that include Northwester University, Cornell, CU, and Texas A\&M.
	
	Dr. Letaief is an internationally recognized leader in wireless communications with research interest in Green communications, Internet of Things, Cloud-RANs, and 5G systems.  In these areas, he has over 560 journal and conference papers and given keynote talks as well as courses all over the world.  He also has 15 patents, including 11 US patents.
	
	Dr. Letaief served as consultants for different organizations including Huawei, ASTRI, ZTE, Nortel, PricewaterhouseCoopers, and Motorola.  He is the founding Editor-in-Chief of the \textsc{IEEE Transactions on Wireless Communications} and has served on the editorial board of other prestigious journals. In addition to his active research and professional activities, Professor Letaief has been a dedicated teacher committed to excellence in teaching and scholarship.  He received the Michael G. Gale Medal for Distinguished Teaching (\emph{Highest university-wide teaching award} and only one recipient/year is honored for his/her contributions).  
	
	He is also the recipient of many other distinguished awards and honors including the 2007 IEEE Joseph LoCicero Publications Exemplary Award, 2009 IEEE Marconi Prize Award in Wireless Communications, 2010 Purdue University Outstanding Electrical and Computer Engineer Award, 2011 IEEE Harold Sobol Award, 2016 IEEE Marconi Prize Paper Award in Wireless Communications, and over 11 IEEE Best Paper Awards.
	
	Dr. Letaief is recognized as a long time volunteer with dedicated service to professional societies and in particular IEEE where he has served in many leadership positions.  These include Treasurer, Vice-President for Conferences and Vice-President for Technical Activities of IEEE Communications Society.   
	
	Dr. Letaief is a Fellow of IEEE and a Fellow of HKIE. He is also recognized by Thomson Reuters as an \emph{ISI Highly Cited Researcher} (which places him among the top 250 preeminent individual researchers in the field of Computer Science and Engineering) and will be the President of IEEE Communications Society in 2018.
\end{IEEEbiography}
	
\end{document}